\documentclass[twocolumn]{aastex62}

\usepackage{color}
\usepackage{soul} 

\newcommand{\lenstool}{{\tt{Lenstool}}}

\newcommand{\Lenstool}{{\tt{Lenstool}}}

\newcommand{\spitzer}{{\it Spitzer}}
\newcommand{\hst}{{\it HST}}
\newcommand{\HST}{{\it HST}}
\newcommand{\HFF}{{\it Hubble Frontier Fields}}
\newcommand{\JWST}{{\it JWST}}

\newcommand{\msun}{M$_{\odot}$}

\newcommand{\clusterfullname}{MACS\,J0417.5$-$1154}
\newcommand{\clustername}{MACS\,J0417}
\newcommand{\zspec}{$z_{spec}$} 
\newcommand{\zphot}{$z_{phot}$} 
\newcommand{\photoz}{photo-$z$}
\newcommand{\modelz}{model-$z$}

\newcommand{\Mtwokpc}{M$_{(200~{\rm kpc})}= 1.78^{+0.01}_{-0.03} \times10^{14}$\msun} 
\newcommand{\MfiveMpc}{M$_{(1.5~{\rm Mpc})}=12.88^{+0.16}_{-0.51} \times10^{14}$\msun} 

\newcommand{\Erad}{$\theta_{E}\simeq$36\arcsec\ } 
\newcommand{\gold}{{\it gold}}
\newcommand{\silver}{{\it silver}}
\newcommand{\bronze}{{\it bronze}}

\received{---}
\revised{---}
\accepted{---}
\submitjournal{ApJ}

\shorttitle{Lensing Analysis of \clustername}
\shortauthors{Mahler et al.}

\begin{document}

\title{RELICS: Strong Lensing Analysis of \clusterfullname\ and Predictions for Observing the Magnified High-Redshift Universe with \JWST}

\correspondingauthor{Guillaume Mahler}
\email{gmahler@umich.edu}

\author[0000-0003-3266-2001]{Guillaume Mahler}
\affil{Department of Astronomy, University of Michigan, 1085 South University Ave, Ann Arbor, MI 48109, USA}

\author[0000-0002-7559-0864]{Keren Sharon}
\affiliation{Department of Astronomy, University of Michigan, 1085 South University Ave, Ann Arbor, MI 48109, USA}

\author{Carter Fox}
\affil{Department of Astronomy, University of Michigan, 1085 South University Ave, Ann Arbor, MI 48109, USA}

\author[0000-0001-7410-7669]{Dan Coe}
\affil{Space Telescope Science Institute, 3700 San Martin Drive, Baltimore, MD 21218, USA}

\author[0000-0003-1974-8732]{Mathilde Jauzac}
\affil{Centre for Extragalactic Astronomy, Department of Physics, Durham University, Durham DH1 3LE, U.K.}
\affil{Institute for Computational Cosmology, Durham University, South Road, Durham DH1 3LE, U.K.}
\affil{Astrophysics and Cosmology Research Unit, School of Mathematical Sciences, University of KwaZulu-Natal, Durban 4041, South Africa}

\author[0000-0002-6338-7295]{Victoria Strait}
\affil{Department of Physics, University of California, Davis, CA 95616, USA}

\author[0000-0002-3398-6916]{Alastair Edge}
\affil{Centre for Extragalactic Astronomy, Department of Physics, Durham University, Durham DH1 3LE, U.K.}

\author[0000-0003-3108-9039]{Ana Acebron}
\affil{Physics Department, Ben-Gurion University of the Negev, P.O. Box 653, Beer-Sheva 84105, Israel}

\author[0000-0002-8144-9285]{Felipe Andrade-Santos}
\affil{Harvard-Smithsonian Center for Astrophysics, 60 Garden Street, Cambridge, MA 02138, USA}

\author[0000-0001-9364-5577]{Roberto J. Avila}
\affil{Space Telescope Science Institute, 3700 San Martin Drive, Baltimore, MD 21218, USA}

\author[0000-0001-5984-0395]{Maru\v{s}a Brada\v{c}}
\affil{Department of Physics, University of California, Davis, CA 95616, USA}

\author[0000-0002-7908-9284]{Larry D. Bradley}
\affil{Space Telescope Science Institute, 3700 San Martin Drive, Baltimore, MD 21218, USA}

\author[0000-0002-3772-0330]{Daniela Carrasco}
\affil{School of Physics, University of Melbourne, VIC 3010, Australia}

\author[0000-0002-8261-9098]{Catherine Cerny}
\affil{Astronomy Department and Institute for Astrophysical Research, Boston University, 725 Commonwealth Avenue, Boston, MA 02215, USA}

\author{Nath\'alia Cibirka}
\affil{Physics Department, Ben-Gurion University of the Negev, P.O. Box 653, Beer-Sheva 84105, Israel}

\author{Nicole G. Czakon}
\affil{Institute of Astronomy and Astrophysics, Academia Sinica, PO Box 23-141, Taipei 10617,Taiwan}

\author[0000-0003-0248-6123]{William A. Dawson}
\affil{Lawrence Livermore National Laboratory, P.O. Box 808 L-210, Livermore, CA, 94551, USA}

\author{Brenda L. Frye}
\affil{Department of Astronomy, Steward Observatory, University of Arizona, 933 North Cherry Avenue, Tucson, AZ, 85721, USA}

\author{Austin T. Hoag}
\affil{Department of Physics, University of California, Davis, CA 95616, USA}

\author{Kuang-Han Huang}
\affil{Department of Physics, University of California, Davis, CA 95616, USA}

\author{Traci L. Johnson}
\affil{Department of Astronomy, University of Michigan, 1085 South University Drive, Ann Arbor, MI 48109, USA}

\author{Christine Jones}
\affil{Harvard-Smithsonian Center for Astrophysics, 60 Garden Street, Cambridge, MA 02138, USA}

\author{Shotaro Kikuchihara}
\affil{Institute for Cosmic Ray Research, The University of Tokyo, 5-1-5 Kashiwanoha, Kashiwa, Chiba 277-8582, Japan}

\author{Daniel Lam}
\affil{Leiden Observatory, Leiden University, NL-2300 RA Leiden, The Netherlands}

\author{Rachael Livermore}
\affil{School of Physics, University of Melbourne, VIC 3010, Australia}
\affil{ARC Centre of Excellence for All Sky Astrophysics in 3 Dimensions (ASTRO 3D), VIC 2010, Australia}

\author{Lorenzo Lovisari}
\affil{Harvard-Smithsonian Center for Astrophysics, 60 Garden Street, Cambridge, MA 02138, USA}

\author{Ramesh Mainali}
\affil{Department of Astronomy, Steward Observatory, University of Arizona, 933 North Cherry Avenue, Tucson, AZ, 85721, USA}


\author{Sara Ogaz}
\affil{Space Telescope Science Institute, 3700 San Martin Drive, Baltimore, MD 21218, USA}


\author{Masami Ouchi}
\affil{Institute for Cosmic Ray Research, The University of Tokyo, 5-1-5 Kashiwanoha, Kashiwa, Chiba 277-8582, Japan}
\affil{Kavli Institute for the Physics and Mathematics of the Universe (Kavli IPMU, WPI), The University of Tokyo, Chiba 277-8582, Japan}





\author{Rachel Paterno-Mahler}
\affil{Department of Astronomy, University of Michigan, 1085 South University Ave, Ann Arbor, MI 48109, USA}

\author{Ian U. Roederer} 
\affil{Department of Astronomy, University of Michigan, 1085 South University Ave, Ann Arbor, MI 48109, USA}
\affil{Joint Institute for Nuclear Astrophysics---Chemical Evolution of the Elements (JINA-CEE), USA}

\author{Russell E. Ryan}
\affil{Space Telescope Science Institute, 3700 San Martin Drive, Baltimore, MD 21218, USA}

\author{Brett Salmon}
\affil{Space Telescope Science Institute, 3700 San Martin Drive, Baltimore, MD 21218, USA}

\author{Irene Sendra-Server}
\affil{Infrared Processing and Analysis Center, California Institute of Technology, MS 100-22, Pasadena, CA 91125}

\author{Daniel P. Stark}
\affil{Department of Astronomy, Steward Observatory, University of Arizona, 933 North Cherry Avenue, Tucson, AZ, 85721, USA}


\author{Sune Toft}
\affil{Cosmic Dawn Center, Niels Bohr Institute, University of Copenhagen, Juliane Maries Vej 30, Kbenhavn, DK-2100, Denmark}

\author{Michele Trenti}
\affil{School of Physics, University of Melbourne, VIC 3010, Australia}
\affil{ARC Centre of Excellence for All Sky Astrophysics in 3 Dimensions (ASTRO 3D), VIC 2010, Australia}

\author{Keiichi Umetsu}
\affil{Institute of Astronomy and Astrophysics, Academia Sinica, PO Box 23-141, Taipei 10617,Taiwan}

\author{Benedetta Vulcani}
\affil{School of Physics, University of Melbourne, VIC 3010, Australia}

\author{Adi Zitrin}
\affil{Physics Department, Ben-Gurion University of the Negev, P.O. Box 653, Beer-Sheva 84105, Israel}








\begin{abstract}
Strong gravitational lensing by clusters of galaxies probes the mass distribution at the core of each cluster and magnifies the universe behind it. 
\clusterfullname\ at $z=0.443$ is one of the most massive clusters known based on weak lensing, X-ray, and Sunyaev-Zel'dovich analyses.
Here we compute a strong lens model of \clustername\
based on {\it Hubble Space Telescope} imaging observations collected, in part, by the Reionization Lensing Cluster Survey (RELICS), and recently reported spectroscopic redshifts from the MUSE instrument on the Very Large Telescope (VLT). We measure an Einstein radius of \Erad at $z=9$ and a mass projected within 200 kpc of \Mtwokpc. Using this model, we measure a ratio between the mass attributed to cluster-member galaxy halos and the main cluster halo of order 1:100.
We assess the probability to detect magnified high-redshift galaxies in the field of this cluster, both for comparison with RELICS {\it HST} results and as a prediction for the \emph{James Webb Space Telescope} (\JWST)\ {Guaranteed Time Observations} upcoming for this cluster. Our lensing analysis indicates that this cluster has similar lensing strength to other clusters in the RELICS program. 
Our lensing analysis predicts a detection of at least a few $z\sim6 - 8$ galaxies behind this cluster, at odds with a recent analysis that yielded no such candidates in this field.
Reliable strong lensing models are crucial for accurately predicting the intrinsic properties of lensed galaxies. As part of the RELICS program, our strong lensing model produced with the \Lenstool\ parametric method is publicly available through the Mikulski Archive for Space Telescopes (MAST).

\end{abstract}



\section{Introduction}
\label{sec:intro}
In our view of the history of the universe, the epoch of reionization remains the least well observed. During the
first billion years, the universe was largely neutral. Half the intergalactic medium (IGM) in the universe was reionized by
 $z = 8 \pm 1$ \citep{Planck2015XLVII}
and nearly completely by $z = 6$. The end of reionization is evidenced by Gunn-Peterson \citep{Gunn1965} troughs (due to absorption by neutral intergalactic hydrogen) observed in $z > 6$ quasar spectra, but not in spectra at $z < 6$ \citep{Becker2001,Becker2015,Djorgovski2001,Fan2006}.
Observing galaxies during the epoch of reionization remains a challenge today. They are much fainter due to their great distance and smaller sizes, and any Lyman-$\alpha$ emission is often scattered or absorbed by the surrounding neutral gas.

Strong lensing magnification by clusters of galaxies offers a privileged view of the high-$z$ universe. Several studies already highlight the high power of gravitational lenses to reveal objects that would have been inaccessible otherwise. Deep observations of Frontier Fields clusters \citep{Lotz2017} were particularly important for probing the faint end of high-redshift luminosity functions and the galaxies most likely responsible for reionization \citep{Atek2015,Livermore2016,Yue2017,Bouwens2017,Ishigaki2018,Bhatawdekar2018,Atek2018}, as well as finding high redshift candidates (e.g. a $z\sim10$ galaxy from  \citealt{Oesch2018}).
The Cluster Lensing And Supernova survey with Hubble (CLASH; \citealt{Postman2012}) yielded $z \sim 6 - 11$ galaxies to be observed more brightly \citep{Hashimoto2018,Bradley2014,Zheng2012,Coe2013}.
Even after these large surveys, many clusters had yet to be observed by the {\it Hubble Space Telescope (HST)} at near-infrared wavelengths (1.0 -- 1.7 $\mu$m) to search for distant galaxies.

\clusterfullname\ (hereafter, \clustername) was discovered by the MAssive Cluster Survey (MACS; \citealt{Ebeling2001}) as part of the ROSAT \citep{Voges1999} catalog of bright sources.
\clustername\ at $z=0.443$ is one of the most X-ray luminous clusters with a luminosity of $2.9 \times 10^{45}$ erg s$^{-1}$ between $0.1-2.4$ keV.
Based on Chandra X-ray observations,
\citet{Mann2012} report that the peak of the X-ray emission is centered on the primary brightest cluster galaxy (BCG) with a slight diffuse emission extended toward the second brightest galaxy in the cluster core. 
\citet{Dwarakanath2011}, \citet{Parekh2017}, and \citet{Sandhu2018} confirm this feature in the radio. \citet{Parekh2017} highlight the similarity in morphology to the clusters Abell~2746 and  1E~0657$-$56 (the ``Bullet cluster''),  
strengthening the hypothesis made by \citet{Mann2012} that \clustername\ is a recent merger, probably oriented along the line of sight, or alternatively, caught close to a turnaround. The merging state of \clustername\ is also confirmed in the analysis of \cite{Pandge2018}. Recently, \citet{Botteon2018} discovered two new cold fronts indicating that substructure dynamics are at play in \clustername. 

\clustername\ was also detected by the \emph{Planck} Early Sunyaev-Zel'dovich (ESZ) catalog \citep{Planck2011VII}, and 
with $M_{500}=(1.23\pm0.05) \times10^{15} M_\odot$
had the fourth highest mass of all 1,094 confirmed clusters with measured redshifts and mass estimates in the \emph{Planck} PSZ2 catalog \citep{Planck2015XXVII}.
Similarly, \clustername\ has the third-highest mass ($M_{1500kpc} = (1.89\pm0.25) \times10^{15} M_\odot$) that was 
measured as a part of a weak lensing analysis of 27 clusters undertaken in the 'Weighing the Giants' census \citep{Applegate2014}.


Based on all of these factors, \clustername\ was included in
the Reionization Lensing Cluster Survey (RELICS).
RELICS is a large {\it Hubble Space telescope} (\HST) Treasury program, GO~14096 (PI: Coe), to observe 46 fields strongly lensed by 41 massive galaxy clusters.
The primary goals of the program are to identify candidates of high-redshift ($6<z<12$) galaxies magnified by the foreground clusters \citep{salmon17,Salmon2018} with photometric redshifts estimated from multi-band imaging with \hst\ and \spitzer\ (PI: Brada\v{c}), and to better constrain luminosity functions at the epoch of reionization. Full details of the project will be described in a forthcoming publication (Coe et al.\ in prep.). Of particular interest is the potential to identify targets to be observed with the {\it James Webb Space Telescope} (\JWST). To support this goal and increase the scientific impact of this program, strong lens models are being computed by the RELICS team \citep{cerny18,Acebron2018a,Acebron2018b,Cibirka2018,Paterno-Mahler2018,Salmon2018} and released to the scientific community via the Mikulski Archive for Space Telescopes (MAST).\footnote{\url{https://archive.stsci.edu/prepds/relics/}}
{The work presented here and the companion paper, \cite{Jauzac2018c}, represent the first public strong lensing analyses on \clusterfullname.}

{
The paper is organised as follows: in Section~2 we give an overview of the data, Section~3 details the strong lensing analysis, and the results are discussed in Section~4, in section~5 we describe predictions for observing the high-redshift universe by current and future facilities, in section~6 we summarize the main results of this work. 

Throughout this paper we adopt a standard $\Lambda$-CDM cosmology with $\Omega_m =0.3$, $\Omega_{\Lambda}=0.7$ and $h=0.7$. All magnitudes are given in the AB system \citep{Oke1974}.
}


\section{Data} \label{sec:style}

\subsection{Imaging}
\subsubsection{\hst}

\clustername\ was first observed by \hst\ in Cycle~16, as part of a snapshot survey of the MACS clusters (SNAP~11103; PI: Ebeling) with the \emph{Wide Field Planetary Camera 2} (WFPC2) in the F606W and F814W bands. Deeper observations with the UVIS instrument on the \emph{Wide Field Camera 3} (WFC3) in the F606W band and with the \emph{Advanced Camera for Survey} (ACS) in the F814W band  were obtained in Cycle~17 as part of the \emph{Chandra} proposal ID \#11800792 (joint with \hst\ GO-12009; PI: von der Linden). It was then observed as part of the RELICS GO program with four filters on the WFC3-IR camera, F160W, F140W, F125W, and F105W; and F435W on ACS.
Our analysis makes use of \hst\ ACS and WFC3 imaging of \clustername, not the original WFPC2 shallow observations.
Table~\ref{tab.hstimaging} lists the dates and exposure times of the \hst\ observations used in this work.

The ACS and WFC3 data were aligned to the same pixel frame and combined using standard procedures as described in \citet{cerny18}. This work made use of images drizzled onto both 30\,mas\,px$^{-1}$ and 60\,mas\,px$^{-1}$, to take advantage of the full resolution capabilities of the WFC3/UVIS and ACS cameras, and proper sampling of the point spread function. We provide fully reduced imaging data as service to the community, and they are publicly available as high-level data products on MAST. 

\begin{deluxetable*}{llcl}
\tablecaption{Details on the observations of MACS\,J0417 taken with the\emph{Hubble Space Telescope}. \label{tab.hstimaging}
 }
\tablecolumns{4}
\tablewidth{0pt}
\tablehead{
\colhead{Camera, filter} &
\colhead{Exp. Time (s)} &
\colhead{UT Date} &
\colhead{Program}
}
\startdata
ACS	F435W	    &	2000.0	&	2016-11-30	&	GO-14096	\\
WFC3/UVIS F606W	&	5364.0	&	2011-01-20	&	GO-12009	\\
WFC3/UVIS F606W	&	1788.0	&	2011-02-28	&	GO-12009	\\
ACS	F814W	    &	1910.0	&	2010-12-10	&	GO-12009	\\
WFC3/IR	F105W	&	705.9	&	2016-12-30	&	GO-14096	\\
WFC3/IR	F105W	&	755.9	&	2017-02-10	&	GO-14096	\\
WFC3/IR	F125W	&	380.9	&	2016-12-30	&	GO-14096	\\
WFC3/IR	F125W	&	355.9	&	2017-02-11	&	GO-14096	\\
WFC3/IR	F140W	&	380.9	&	2016-12-30	&	GO-14096	\\
WFC3/IR	F140W	&	355.9	&	2017-02-10	&	GO-14096	\\
WFC3/IR	F160W	&	1005.9	&	2016-12-30	&	GO-14096	\\
WFC3/IR	F160W	&	1005.9	&	2017-02-11	&	GO-14096	\\
\enddata
\end{deluxetable*}

\subsubsection{\spitzer}

The \emph{Infrared Array Camera} (IRAC) on board the \emph{Spitzer} space telescope imaged MACSJ\,0417 as part of the S-RELICS programme (\emph{Spitzer}-RELICS, PI: Brada{\v c}, PI: Soifer). Observations reach 13 hours of total exposure time in each of IRAC channels 1 and 2 (3.6\,$\mu$m and 4.5\,$\mu$m). The data reduction will be described in detail in Strait et al.\ (in prep.); to create the mosaic images we use the \textsc{MOsaicker and Point source EXtractor} (\textsc{mopex}\footnote{http://irsa.ipac.caltech.edu/data/SPITZER/docs/\\dataanalysistools/tools/mopex/}) and largely follow the process described in the IRAC Cookbook\footnote{http://irsa.ipac.caltech.edu/data/SPITZER/docs/\\dataanalysistools/cookbook/} for the COSMOS medium-deep data.

The intra-cluster light subtraction and flux extraction are done using \textsc{t-phot} \citep{merl15}, designed to perform PSF-matched, prior-based, multi-wavelength photometry as described in \cite{merl15,merl16}. This is done by convolving cutouts from a high resolution image (in this case, F160W) using a low resolution PSF transformation kernel that matches the F160W resolution to the IRAC (low-resolution) image. \textsc{t-phot} then fits a template to each source detected in F160W to best match the pixel values in the IRAC image. The IRAC fluxes are then combined with \emph{HST} fluxes in catalogs.

\subsection{Spectroscopy}
\subsubsection{LDSS3}
We obtained multislit spectroscopy of \clustername\ with the upgraded Low Dispersion Survey Spectrograph (LDSS3-C)\footnote{\url{http://www.lco.cl/telescopes-information/magellan/ instruments/ldss-3}} on the Magellan Clay telescope, on 2017 July 27 using University of Michigan allocation (PI: Sharon). Two multislit masks were designed, with $1\farcs0$ slits placed on multiple images of lensed galaxies at the highest priority, and the rest of the mask filled with background sources and cluster-member galaxies. Due to weather conditions, only one of the masks was observed, with three exposures of 1200\,seconds each. The seeing ranged between $0\farcs5-0\farcs7$, with some clouds present during the observation. The data were obtained with the VPH-ALL grism ($4250\,\textrm{\AA} < \lambda < 10000\,\textrm{\AA}$) with spectral resolution R$=$450-1100 across the wavelength range. The spectroscopic data were reduced using the standard procedures using the COSMOS data reduction package \citep{Dressler2011,Oemler2017}. We measured a spectroscopic redshift of $z_{spec} = 0.871$ for image 1.3 
($\alpha$ = 04:17:33.70, $\delta$ = -11:54:39.70), based on [OII] $\lambda$ 3728 and H$\beta$ line emission.
 The data yielded a spectroscopic redshift for another background source ($\alpha = $4:17:35.942, $\delta = -$11:54:59.29), at \zspec$=1.046$, from [OII]\,$\lambda$3728, [OIII] $\lambda$4959,5007; however, this source is not multiply-imaged and was not used as a constraint in the lens model (see Section~\ref{sec:lensing}). 

A full description of the RELICS Magellan/LDSS3 follow-up results will be presented in a future paper (Mainali et al.\ in prep).

\subsubsection{MUSE}
The field was observed with the Multi Unit Spectrographic Explorer (MUSE; \citealt{Bacon2010}) on 2017 December 12.
The MUSE exposure was 3$\times$970\,s, or 2910\,s in total, and was
taken as part of ESO project 0100.A-0792(A) (PI: Edge). The data were reduced and spectra extracted as explained in the companion paper \cite{Jauzac2018c}. The MUSE field of view, $1\arcmin \times1 \arcmin$, is approximately centered on the BCG, and does not cover the full extent of the \hst\ field of view. The MUSE spectral resolution is R$=$1750--3750 across the wavelength range 4800 -- 9300$\,\textrm{\AA}$.

This work makes use of the spectroscopic redshifts measured for lensed galaxies reported in the companion paper by \cite{Jauzac2018c} (Table~\ref{tab:arcs}). The MUSE observation confirms the redshift that was obtained with LDSS3 for image 1.3, \zspec$=0.871$, and spectroscopically confirms images 1.1 and 1.2 as counter images of the same system. Moreover, it reveals [OII]\,$\lambda$3728 emission from a fourth image at the same redshift, buried in the light of the BCG. This fourth image is likely not a complete image, therefore we did not use it as a constraint to model the cluster
{The redshift of system~2 and system~3 are both measured at \zspec=$1.046$. The two systems correspond to two different galaxies separated by $\sim 140$ kpc in the source plane according to our modeling. 

For image 4.2 and 4.1, the MUSE data are consistent with a low-confidence redshift of $z$=3.10. Due to the low confidence of this measurement, we do not use it as a constraint.}
A full description of the data and results related to other objects in the field are given in our companion paper by \cite{Jauzac2018c} 

\section{Gravitational Lensing Analysis} 
\label{sec:lensing}

\subsection{Methodology}
\label{sec:methodology}
The lens model of \clustername\ was computed using the public software \lenstool\ \citep{Jullo2007}, which is a parametric lens modeling algorithm that employs Markov Chain Monte Carlo (MCMC) analysis to explore the parameter space and identify the best-fit solution. The lens plane is modeled as a linear combination of several mass halos, each parameterized as a pseudo isothermal ellipsoidal mass distribution (PIEMD or dPIE; \citealt{Eliasdottir2007}) with seven parameters: position $x$, $y$; ellipticity $\varepsilon$; position angle $\theta$; core radius $r_c$; cut radius $r_{cut}$; and normalization $\sigma_0$. 
The two radii parameters, $r_c$ and $r_{cut}$, define the region $r_c \lesssim r \lesssim r_{cut}$ in which the mass profile is isothermal; the mass density transitions smoothly, but drops rapidly beyond $r_{cut}$.
The cluster mass distribution is typically dominated by cluster-scale and group-scale halos, whose parameters are set free. Galaxy-scale halos are placed at the observed positions of cluster-member galaxies, with positional parameters ($x$, $y$, $\theta$, $\varepsilon$) fixed at the observed values of their light distribution as measured with SExtractor \citep{SEx} and the other parameters scaled with the luminosity of the galaxy in F814W, following scaling relations as described in \citet{Limousin2005}. The normalization parameter of the scaling relation, $\sigma_0^*$, is a free parameter. The slope parameters and normalization of the three brightest galaxies are solved individually, and those are decoupled from the scaling relations of the other cluster members. The BCG is clearly bluer than the cluster red sequence  due to ongoing star formation \citep{Green2016}
and therefore is not expected to follow the same scaling relation \citep{Postman2012}. The other two galaxies dominate the subgroups at the north of the field of view, and by leaving their parameters free we allow for a larger contribution of underlying dark matter halo in this region. 

An alternative approach would be to model these two galaxies separately
and adding two other group-scale halos to model their dark matter component, allowing flexibility in the position of the underlying total potential, as it is done in our companion paper by \cite{Jauzac2018c}

Following the red sequence technique of \cite{Gladders2000}, we select cluster members from a color-magnitude diagram using F606W-F814W vs F814W, bracketing appropriately the 4000\,\AA\ break, which is a typical feature observed in elliptical galaxies. We selected galaxies down to 24 magnitude, which corresponds approximately to 0.01\,L* at redshift 0.44 resulting in 177 galaxies identified as cluster members in total. The magnitude is referred to the SExtractor parameters {\rm MAG\_AUTO} as defined in Sextractor \citep{SEx}.

The lens model is constrained with sets of multiple images, identified in the \hst\ imaging data and classified as described below. The position of each image is used as a constraint. Where substructure is clearly identified and can be robustly matched between images, we use multiple emission knots in each image, which indirectly constrains the relative magnification between images. We refrain from over-weighting systems by limiting the number of emission knots used in any single image to four. 

Where known, spectroscopic redshifts are used as fixed redshift constraints. These are available for systems 1, 2, and 3. Most of the other systems have photometric redshifts from the RELICS analysis. However, 
following \cite{cerny18} and \cite{Johnson2016}, who studied the effects of redshift accuracy on the lens model, the redshifts of systems with no spectroscopic redshifts (\zspec) are left as free parameters with broad limits, to avoid biases due to photometric redshifts (\zphot) outliers. We check the model-predicted source redshifts against the photometric redshift in Section~\ref{sec:photoz} 
as an independent confirmation that the model is not converging onto a completely wrong solution  \citep[see discussion in][]{cerny18}.   

\subsection{Lensing Constraints}
\label{sec:cstr}
\begin{figure*}
\begin{center}
 \includegraphics[width=\textwidth]{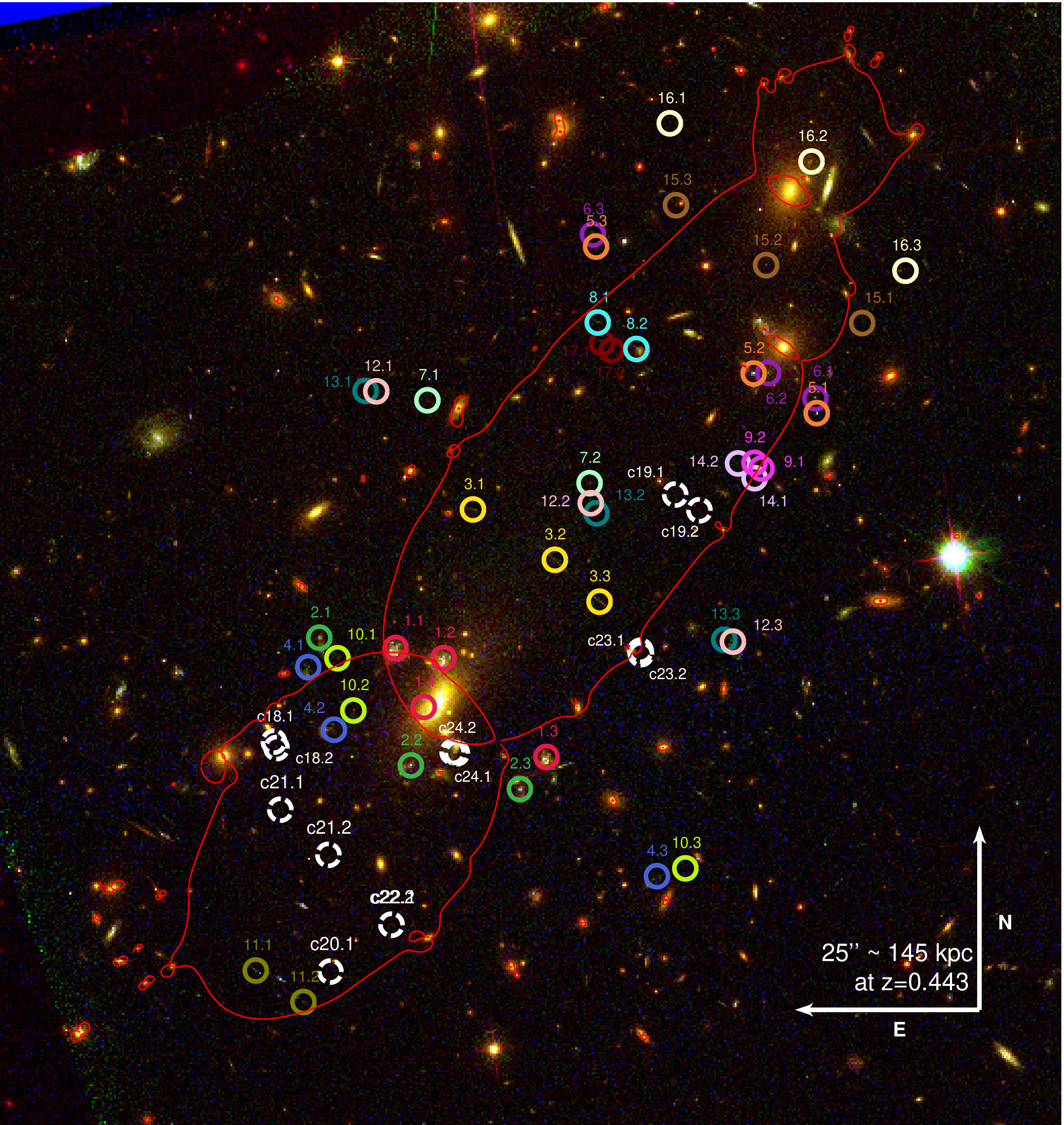}
\caption{Composite color image of \clustername\ created from \hst\ imaging in ACS F814W (red), WFC3/UVIS F606W (green), and ACS F435W (blue). Secure multiply-imaged galaxies (\gold, \silver, and \bronze) are labeled with colored circles, color-coded by system. The white dashed circles label candidate images that were not used as constraints. The red line marks the location of the critical curve for a source at $z=9$.}
\label{fig:img-mul}
\end{center}
\end{figure*}

We identify 57 images of 17 systems that are used as constraints and 7 candidates of strongly-lensed images. Following the \HFF\ ranking process, we classify the observed lensed images into three categories: \gold, \silver, and \bronze. The \gold\ category includes robustly-identified multiply-imaged systems with a measured spectroscopic redshift; three systems fall in this category. The \silver\ classification is given to multiply-imaged systems that are reliably identified as such by morphology, surface brightness, and lensing symmetry; 12 systems fall in this category. 
Images that have less robust identification, or would not be identified as counter images without an accurate lens model, were put in the \bronze\ category and not used as constraints in our fiducial (\silver) model. All systems are shown in Figure \ref{fig:img-mul}, and their coordinates, redshifts, and ranking, are tabulated in Table~\ref{tab:arcs}. 
We note that system~4 has a possible redshift of 3.1 from MUSE, however, it is based on low-confidence features. We choose to not include the redshift as a constraint in the model, as if it is incorrect the redshift might bias the model as was shown by, e.g. \cite{Jauzac2015}, \cite{Johnson2016}, \cite{cerny18}, and \cite{Remolina2018}. 

We identify several other strong lensing features in the field, which, at the depth of the data in hand, are not deemed reliable enough to be used as constraints. We list these candidates in this paper for completeness. 
All the candidates are presented in Figure~\ref{fig:img-mul}, and their coordinates 
are tabulated in Table~\ref{tab:candidates} in the Appendix.

\subsection{Mass model components}
\label{sec:DM}
As described in Section~\ref{sec:methodology}, and typical for parametric lens modeling algorithms, the lens plane is described by a combination of several dark matter (DM) halos whose parameters are allowed to vary, with contribution from galaxy-scale halos that follow scaling relations. 
The lens model of \clustername\ includes four ``free'' DM halos, all parameterized as PIEMDs. The dominant component is a cluster-scale halo, whose parameters are all allowed to vary, with the exception of the truncation radius $r_{cut}$ that extends beyond the strong lensing regime and cannot be constrained by the strong lensing evidence. 
Three other halos are placed on the three brightest cluster galaxies, with positional parameters ($x$, $y$, $\varepsilon$, $\theta$) following their light distribution, and the other parameters set free. 
We emphasize that the halos placed on these galaxies are not to be considered strictly galaxy halos. The model cannot disentangle the dark matter halo in which the galaxy is embedded from the underlying dark matter halo of the cluster or group. 

\startlongtable
\begin{deluxetable*}{llllllllc} 
\tablecolumns{7} 
\tablecaption{List of lensing constraints. R.A. and Decl. refer to right ascension and declination of the constraints position. $z_{spec}$ refers to the spectroscopic constraints when available; references for the spectroscopic redshifts are given in the table footnotes. $z_{model}$ indicates the best-fit redshift estimates resulting from the ``silver'' and ``bronze'' lens models with their respective statistical uncertainties. 
The rms is the difference between the observed position of a multiple image and the predicted position from the barycenter of our best-fit model in the image plane given in arcseconds. 
The classification scheme is discussed in Sect.~\ref{sec:cstr}. \label{tab:arcs}} 
\tablehead{\colhead{ID} &
            \colhead{R.A.}    & 
            \colhead{Decl.}    & 
            \colhead{$z_{spec}$}     & 
            \colhead{$z_{model}$}       & 
            \colhead{$rms$ ($''$)}       & 
           \colhead{$z_{model}$}       &           
           \colhead{$rms$ ($''$)}       & 
            \colhead{Classification}       \\[-8pt]
            \colhead{} &
            \colhead{J2000}     & 
            \colhead{J2000}    & 
            \colhead{}       & 
            \colhead{\silver\ }       & 
            \colhead{\silver\ }       & 
            \colhead{\bronze\ }       & 
            \colhead{\bronze\ }       & 
            \colhead{}             }
\startdata 
1a.1 & 64.396158 & -11.906760 & 0.8710$^{\rm a}$ &\nodata  & 0.07 &\nodata & 0.10 & \gold\ \\ 
1a.2 & 64.394310 & -11.907136 &   & & 0.16 & & 0.22 &  \\ 
1a.3 & 64.390348 & -11.910864 &   & & 0.09 & & 0.09 &\\ 
1b.1 & 64.396081 & -11.907255 &   & & 0.32 & & 0.36 &\\ 
1b.2 & 64.394729 & -11.907583 &   & & 0.43 & & 0.53 &\\ 
1b.3 & 64.390299 & -11.911274 &   & & 0.09 & & 0.13 &\\ 
1c.2 & 64.394371 & -11.907409 &   & & 0.26 & & 0.35 &\\ 
1c.1 & 64.396364 & -11.906983 &   & & 0.13 & & 0.17 &\\ 
1c.3 & 64.390488 & -11.911052 &   & & 0.10 & & 0.12 &\\ 
\hline
2a.1 & 64.399096 & -11.906369 & 1.0460$^{\rm b}$ &\nodata & 0.41 &\nodata & 0.47 &\gold\ \\ 
2a.2 & 64.395567 & -11.911182 &  & & 0.42 && 0.45 & \\ 
2a.3 & 64.391371 & -11.912074 &  & & 0.18 && 0.24 & \\ 
2b.1 & 64.399000 & -11.906633 &  & & 0.50 && 0.57 & \\ 
2b.2 & 64.395821 & -11.911226 &  & & 0.47 && 0.48 &\\ 
2b.3 & 64.391262 & -11.912324 &  & & 0.24 && 0.30 & \\ 
2c.1 & 64.399004 & -11.906855 &  & & 0.49 && 0.56 & \\ 
2c.2 & 64.395954 & -11.911299 &  & & 0.48 && 0.49 & \\ 
2c.3 & 64.391300 & -11.912493 &   & & 0.24 && 0.30 & \\ 
\hline
3.1 & 64.393180 & -11.901537 & 1.0460$^{\rm b}$  &\nodata  & 0.72 &\nodata& 0.59 & \gold\ \\ 
3.2 & 64.390026 & -11.903434 &  &  & 0.81 && 0.89 & \\ 
3.3 & 64.388304 & -11.905013 &  &  & 0.30 & & 0.56 &\\ 
\hline
4.1 & 64.399521 & -11.907479 & \nodata   & $2.26^{+0.08}_{-0.08}$ & 0.26 & $2.33^{+0.07}_{-0.07}$ & 0.50 &\silver\  \\ 
4.2 & 64.398529 & -11.909839 &   & & 0.65 & & 0.47 &\\ 
4.3 & 64.386095 & -11.915359 &   & & 0.60 & & 0.34 & \\ 
\hline
5.1 & 64.379941 & -11.897906 & \nodata   & $2.27^{+0.11}_{-0.14}$ & 0.24 & $2.25^{+0.09}_{-0.07}$ & 0.20 & \silver\  \\ 
5.2 & 64.382370 & -11.896413 &  &  &  0.27 & & 0.25 & \\ 
5.3 & 64.388438 & -11.891630 &  &  &  0.51 & & 0.49 & \\ 
\hline
6.1 & 64.379991 & -11.897349 & \nodata   & $2.34^{+0.11}_{-0.17}$ & 0.21  & $2.27^{+0.09}_{-0.06}$ & 0.24 & \silver\  \\ 
6.2 & 64.381808 & -11.896390 &  &  &  0.18 &  & 0.15 &\\ 
6.3 & 64.388558 & -11.891170 &  &  &  0.46 & & 0.42 & \\ 
\hline
7.1 & 64.394933 & -11.897423 & \nodata   & $^{\rm d}2.09^{+0.05}_{-0.05}$ & $^{\rm d}$0.18  & $2.09^{+0.12}_{-0.08}$ & 0.32 &\bronze\ \\ 
7.2 & 64.388688 & -11.900546 &  &  & $^{\rm d}$0.26 &  & 0.28 &\\ 
\hline
8.1 & 64.388372 & -11.894492 & \nodata    & $2.39^{+0.14}_{-0.12}$ & 0.12 & $2.35^{+0.12}_{-0.09}$ & 0.06 &\silver\  \\ 
8.2 & 64.386885 & -11.895489 &  &  & 0.14 &  & 0.04 &\\ 
\hline
9.1 & 64.382068 & -11.899994 & \nodata    & $5.97^{+0.01}_{-0.20}$ & 0.35  &  $5.52^{+0.10}_{-0.31}$ & 0.43 & \silver\  \\ 
9.2 & 64.382338 & -11.899779 &  &  & 0.22 &  & 0.24 &\\ 
\hline
10.1 & 64.398397 & -11.907143 & \nodata   & $2.02^{+0.14}_{-0.10}$ & 0.28  & $2.33^{+0.07}_{-0.09}$ & 0.43 &\silver\ \\ 
10.2 & 64.397785 & -11.909114 &  &  & 0.37  &  & 0.76 & \\ 
10.3 & 64.385000 & -11.915063 &  & $^{\rm d}2.34^{+0.05}_{-0.04}$ & $^{\rm d}$0.30  & & 0.33 & \bronze$^c$ \\ 
\hline
11.1 & 64.401544 & -11.918912 & \nodata &  $3.47^{+0.36}_{-0.31}$ & 0.18 & $3.18^{+0.25}_{-0.11}$ & 0.09 &\silver\ \\ 
11.2 & 64.399708 & -11.920099 & &    & 0.30 &  & 0.41 & \\ 
\hline
12.1 & 64.396902 & -11.897085 & \nodata   &  $2.84^{+0.13}_{-0.13}$ & 0.42 & $2.81^{+0.16}_{-0.14}$ & 0.34 & \silver\  \\ 
12.2 & 64.388640 & -11.901300 &  &  &0.77 &  & 0.62 &\\ 
12.3 & 64.383172 & -11.906519 &  &   &0.26 &  & 0.19 &\\ 
\hline
13.1 & 64.397312 & -11.897068 & \nodata    & $2.89^{+0.15}_{-0.13}$ & 0.36 & $2.85^{+0.14}_{-0.17}$ & 0.32 & \silver\  \\ 
13.2 & 64.388420 & -11.901684 &  &   &0.73 &  &0.58 &\\ 
13.3 & 64.383499 & -11.906446 &  &  &0.28 &  & 0.17 &\\ 
\hline
14.1 & 64.382335 & -11.900359 & \nodata    & $^{\rm d}4.40^{+0.43}_{-0.21}$ & $^{\rm d}$0.03  & $4.43^{+0.29}_{-0.39}$ & 0.10 &\bronze\ \\ 
14.2 & 64.382972 & -11.899802 &  &  & $^{\rm d}$0.03 &  & 0.12 & \\ 
\hline
15.1 & 64.378193 & -11.894510 & \nodata   & $2.11^{+0.16}_{-0.16}$ & 0.28 & $2.09^{+0.09}_{-0.08}$ & 0.15 &   \silver\  \\ 
15.2 & 64.381890 & -11.892331 &  &  & 0.29 &  & 0.20 &\\ 
15.3 & 64.385361 & -11.890071 &  &  & 0.15 & & 0.04 & \\ 
\hline
16.1 & 64.385599 & -11.886984 & \nodata    & $4.50^{+1.91}_{-0.96}$ & 0.16 & $4.66^{+0.58}_{-0.33}$ & 0.16 & \silver\  \\ 
16.2 & 64.380143 & -11.888425 &  &  & 0.31 &  & 0.26 &\\ 
16.3 & 64.376525 & -11.892540 &  &  & 0.02 &  & 0.30 &\\ 
\hline
17.1 & 64.388212 & -11.895269 & \nodata    & $2.30^{+0.10}_{-0.11}$ & 0.21 & $2.16^{+0.14}_{-0.06}$ & 0.10 &\silver\  \\ 
17.2 & 64.387833 & -11.895536 &  &  & 0.24 &  & 0.11 & \\ 
\enddata 
\tablecomments{$^{\rm a}$ Spectroscopic redshift from Magellan / LDSS3 (this work) and confirmed by MUSE in \citet{Jauzac2018c}.\\[1pt]
$^{\rm b}$ Spectroscopic redshifts from MUSE presented in \citet{Jauzac2018c}. Sources 2 and 3 are at the same redshift. These galaxies are separated by $\sim 140$\,kpc in the source plane. \\[1pt]
$^{\rm c}$ In system 10, images 10.1 and 10.2 are classified as \silver\ and 10.3 is classified as \bronze. Image 10.3 was therefore not included in the ``silver'' model. \\[1pt]
$^{\rm d}$ These redshifts and rms values are computed using the best fit model computed with \silver\ constraints fixed, while only the redshifts of those systems are being optimized.}
\medskip
\end{deluxetable*} 

Table~\ref{tab:fit_param} lists the best fit parameters of each halo, for several lens models. The ``Silver'' model uses as constraints the \gold\ and \silver\ arcs; The ``Bronze'' model uses \gold, \silver, and \bronze\ constraints. We describe the third test model, labeled ``Bridge'', below. 

As can be visually gleaned from the distribution of galaxies (Fig.~\ref{fig:img-mul}), the cluster core is fairly elongated, with the second and third brightest galaxies significantly separated in projection from the BCG. 
In the X-ray, \citep{Mann2012,Parekh2017} report extended emission elongated in the SE--NW direction. 
We, therefore, compute an additional lens model that includes a fifth PIEMD DM halo, forming a mass ``bridge'' between the central and NW components. We test this hypothesis using the \gold+\silver\ list of constraints. 
{The fifth halo is free to vary between the BCG and the NW component. The core radius of the potential is intentionally free to vary up to a high value (300\,kpc) to allow a possible flat profile. The cut radius is fixed to a 1.5\,Mpc as the main DM halo potential}

We quantitatively compare the quality of the three lens models using two criteria. The first one is the rms, which describes how well the model reproduces the image-plane positions of the constraints. The second one is the Bayesian Information Criterion (BIC, introduced by \citealt{Schwarz1978}), which is a statistical measurement based on the model Likelihood $\mathcal{L}$, penalized by the number of free parameters $k$ and the number of constraints $n$: 
\begin{equation}
{\rm BIC} = -2 \times \log(\mathcal{L}) + k \times \log(n),
\end{equation}
The rms gives a good indication of the global distance between the predicted image positions compared to the observed ones, thus for a fixed number of constraints a low rms generally implies a better model. The BIC quantifies an improvement in the model likelihood while taking into account a possible difference in the number of parameters and/or constraints between models. Thus a favorable model will be one with the best likelihood while keeping the lowest BIC value possible. Such criteria were used in previous analyses \citep{Lagattuta2017,Mahler2018,Jauzac2018a} to compare different variation models for a single cluster.
The rms of the ``Bridge'' model is slightly better ($0\farcs36$) compared to the fiducial model ($0\farcs37$). However, the BIC shows an opposite trend when comparing the two models.
We interpret a higher BIC value for the ``Bridge'' model as an over-fit of the model compared to a model without the bridge. In other words, the model does not improve enough to justify the addition of new parameters.
Similar statistical analyses were made in other studies, e.g.: using a discrimination by the evidence \citep{Limousin2010}, other Likelihood penalization: Akaike Information Criterion \citep{Acebron2017} or a combination of a large number of indicators \citep{Jauzac2018a}.

\begin{figure*}
\begin{center}
 \includegraphics[width=0.8\textwidth]{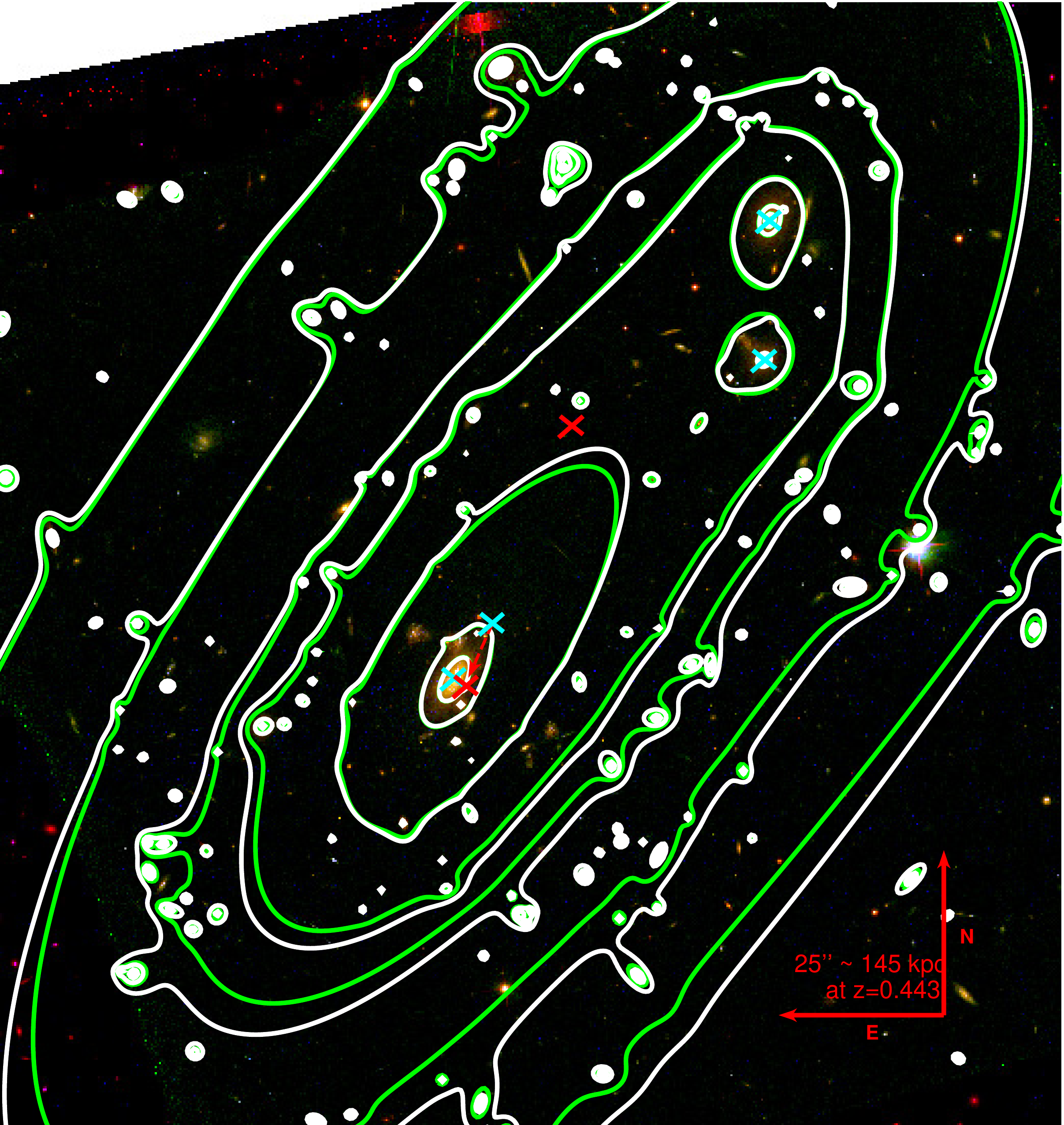}
\caption{
The cyan crosses show the position of all the individual DM potentials for our fiducial model.
The top red cross shows the position of the center of the bridge potential. The red arrow indicates the shifted location of the main DM halo located at the red cross.
The contour at $1.5 \times 10^{9}$\msun kpc$^{-2}$ guides the eye to the apparent comet-like profile as seen in the X-ray luminosity distribution reported by previous studies \citep{Ebeling2014,Parekh2017,Sandhu2018}. A direct comparison between the DM and X-ray light distributions is shown and discussed in \citet{Jauzac2018c}. 
The projected mass density distributions are similar between the models in areas North of the BCG, and their contours are virtually indistinguishable around the BCG where the mass distribution is well constrained. The main differences between the models appear in the South-East, due to the lack of constraints in that side of the cluster (see sect \ref{sec:DM} for more details).
}

\label{fig:mass_contours}
\end{center}
\end{figure*}

{We compare the mass distribution between the models and plot their mass contours in Fig.~\ref{fig:mass_contours}.
The difference between the two models is most notable the South-East region of the cluster. While the BCG area is well-constrained by systems surrounding the BCG, there is only one system with two images farther out. A confirmation of some of the lensed galaxy candidates with deeper observations would better constrain this region.}

\section{Discussion of lens model results}
\label{sec:analysis}
The spectroscopic capabilities of MUSE allow us to detect a central image for system~1 buried in the light of the BCG. Our model predicts a radial pair at this location, however, only a single peak of emission is visible. We interpret that as the likely result of the source-plane caustic bisecting the galaxy in the source plane, resulting in a merging pair configuration where only a small fraction of the source galaxy is lensed into these positions. A more detailed analysis of the lensing configuration of this galaxy is presented in the companion paper by \cite{Jauzac2018c}.

\begin{figure*}
\begin{center}
    \includegraphics[width=\textwidth]{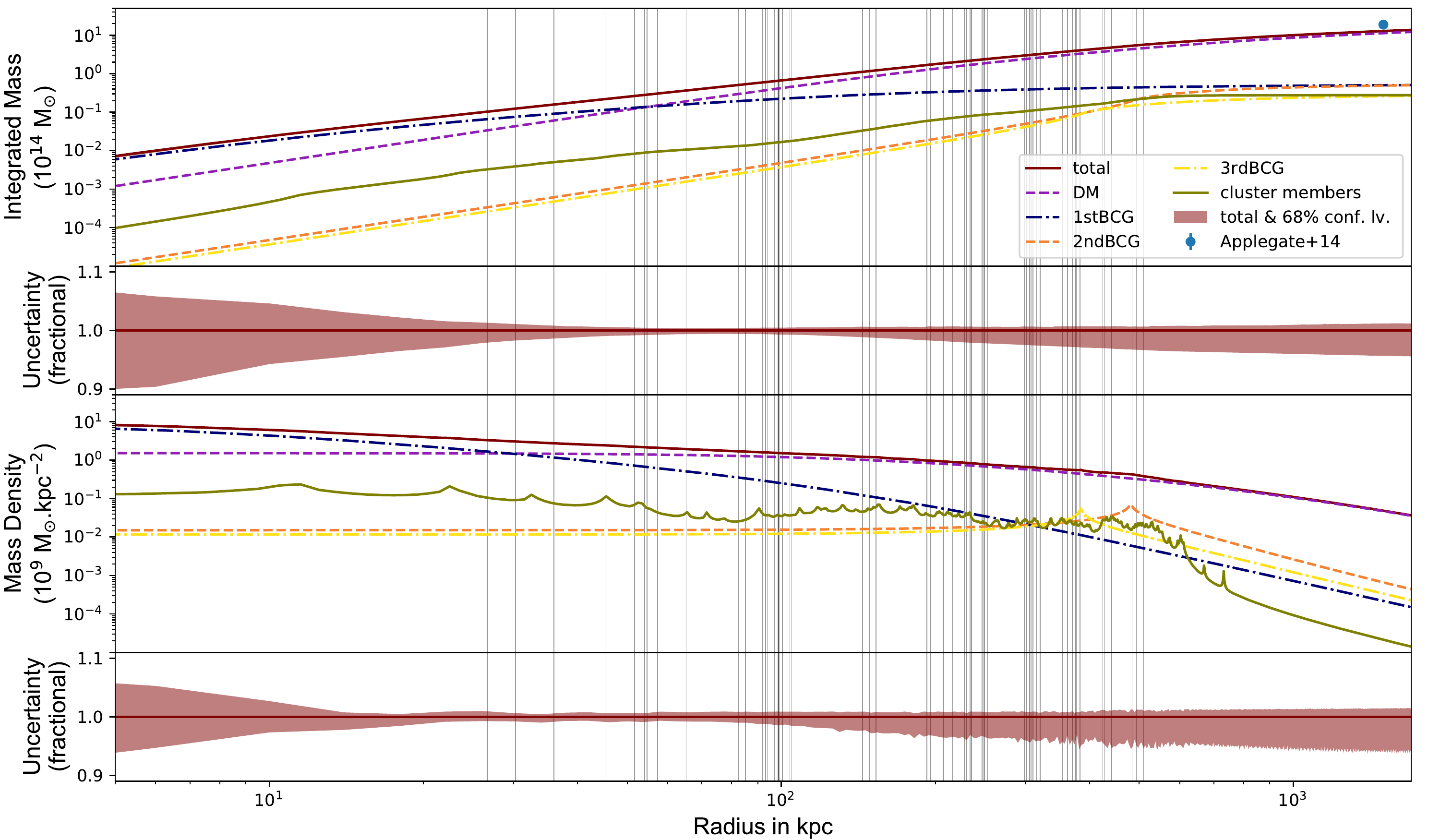}
        \caption{{\it Top}: Integrated mass profiles within a circular aperture centered on the BCG. Our parametric approach enables us to separate the different components of our mass profile. The profile labeled \emph{total} represents our best fiducial model (i.e. using \gold\ and \silver\ constraints). The profile labeled \emph{DM} represents the cluster-scale dark matter halo component (see Sect.~\ref{sec:DM}). The profiles labeled \emph{1stBCG}, \emph{2ndBCG}, and \emph{3rdBCG} show the contribution of the three dark matter potentials placed at the locations of the three brightest galaxies of the cluster. The profile labeled \emph{cluster members} represents the contribution of all cluster member galaxies excluding the brightest three. We find a ratio between the main dark matter halo and the clusters members dark matter halo of about 100:1.
        Strong lensing constraints are plotted as vertical gray lines at their projected distance from the BCG. This is done to highlight where lensing constraints are available.
        In the regions where multiple images are not identified, the mass profile is an extrapolation. Weak lensing mass measurement from \cite{Applegate2014} is plotted as a blue symbol.
        {\it Bottom}: Mass density profiles as a function of the distance to the cluster center. The color coding follows the one from the top panel.
        {The dark-red shaded areas show the $68\%$-confidence interval statistical uncertainty for the total mass profile, with the fractional error shown below each panel. We note that the small statistical uncertainties derived from the modeling underestimate the true error, which is driven by systematic uncertainties.}
        }
    \label{fig:mass_profile}
    \end{center}
    
\end{figure*}

We report an effective Einstein radius of \Erad for a source at $z=9$. The effective Einstein radius is the radius of a circle with the same area as an ellipse fitted to the critical curve. 
We measure a total projected mass of \Mtwokpc\ within 200 kpc.
Figure~\ref{fig:mass_profile} shows the radial mass profile centered on the BCG. Using the capability of our parametric approach we compute the mass profile of five different components of our cluster model: the main cluster-scale dark matter halo, the halos centered on the three brightest cluster galaxies, and the mass distribution of all the other galaxies, which follow a mass-to-light relation.

We qualitatively report a mass ratio of order 100:1 between the main cluster halo and the mass associated with the light of cluster elliptical galaxies, excluding the three brightest galaxies (dark green and magenta lines in  Figure~\ref{fig:mass_profile}). This is consistent with the relative mass to light ratio of rich clusters of about $10^{14}$\msun\ as reported in \citet{Girardi2002}. 
One will see that this qualitative result has no uncertainties attached, since the statistical uncertainties of the mass profile in Section \ref{fig:mass_profile} are likely underestimating the true uncertainty 
due to modeling assumptions (e.g., \citealt{Meneghetti2017}  and structure along the line of sight \citep{Chirivi2018}.

\subsection{Photometric Redshifts}
\label{sec:photoz}

\begin{figure*}
\begin{center}
\includegraphics[trim={4cm 1cm 2cm 2cm},width=\textwidth]{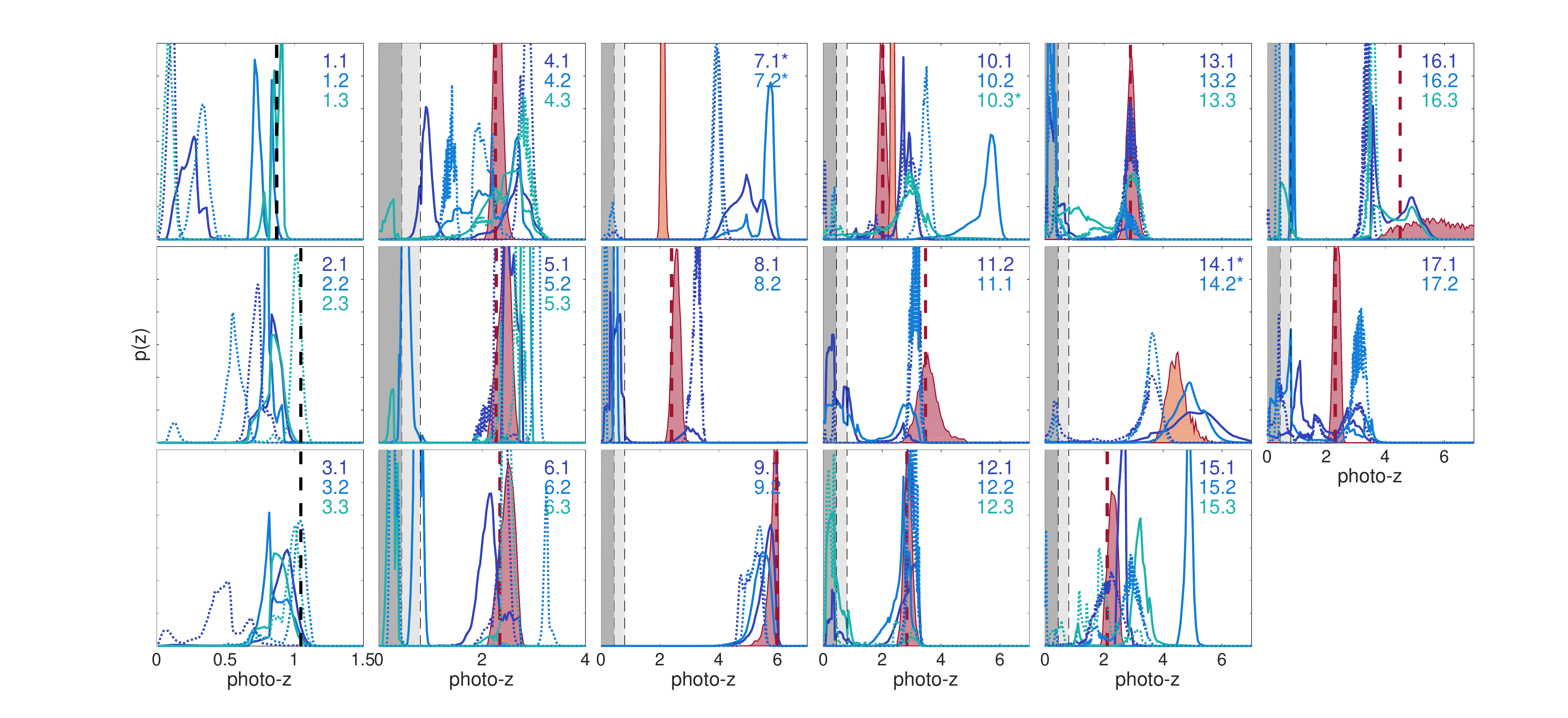}
\caption{Redshift probability distribution functions (PDFs) of the multiply-imaged galaxies used as constraints in the lensing analysis.
The blue lines represent photometric redshift PDF estimates from \textsc{bpz} using the seven \HST\ bands (dotted lines) and from \textsc{eazy} using the seven \HST\ and two \spitzer\ bands (solid lines). The red shaded distributions are our lens model estimates based on MCMC sampling of the parameter space. The red vertical dashed lines show the best fit value, model-$z$, for each system. The light orange shaded areas show predictions from the fiducial (``silver'') lens model for multiple images not included in the ``silver'' set: the \bronze\ systems, 7 and 14, and system 10 with its third counter image 10.3 is included. 
Systems 1, 2, and 3 have a measured spectroscopic redshift shown as vertical black dashed lines. 
The dark gray shaded area marks the redshift range in front of the cluster ($z<0.443$). The light gray shaded area marks the redshift range $0.443<z<0.8$, for which sources 4 -- 17 would not be strongly lensed. 
The numbers in each panel correspond to the multiple image identification numbers as reported in Fig.~\ref{fig:img-mul} and Table~\ref{tab:arcs}. An asterisk marks the \bronze\ galaxies. See Sect.~\ref{sec:analysis} for more detail. }
\label{fig:modelz-photoz}
\end{center}
\end{figure*}

{The lens modeling procedure let the redshift of multiply-imaged systems with no spectroscopic confirmation free to vary. It explores the parameter space to find the most likely redshift (\modelz) of each system. Generally, we find that the redshifts predicted by the ``Silver'' model are in agreement with those predicted by the ``Bronze'' model.
However, a comparison between the lens model-predicted redshifts (\modelz) and photometric redshift (\photoz) estimates can be used for a qualitative assessment of the validity of the lens model. 

The RELICS program delivered photometric redshift catalogs using \textsc{bpz} \citep{Benitez00,Coe06}
based on \hst\ photometry measured in ACS and WFC3 images.
We compare our \modelz\ results against photometric redshifts from the public catalog, as well as against a photometric redshift analysis that supplements the \hst\ data with \spitzer\ photometry and uses a different algorithm, \textsc{eazy} \citep{Brammer2008}. A thorough description of the \hst+\spitzer\ \zphot analysis will be provided in a forthcoming paper (Strait et al.\ in prep).

The multiplicity (i.e., having multiple images for each lensed source) provides an additional means to test the robustness of the photometric redshifts of the lensed galaxies in this field. 
We note that in some cases, the \photoz\ measured for the different multiple images of the same source can disagree. That happens even for systems where the visual identification of the multiple images is entirely unambiguous. This discrepancy could be due to contamination from nearby sources (usually cluster members); variations in SExtractor's detection, deblending, and segmentation for each multiple image; and \photoz\ degeneracies, especially when the lensed image is very faint (mag $> 28$). 
Structure along the line of sight could potentially increase the uncertainty of the \modelz, and contribute to a discrepancy between \modelz\ and \photoz. However, this effect is likely not significant, and not the main source of discrepancy \citep{Chirivi2018}.

For these reasons, it is instructive to examine the entire probability distribution function (PDF) of the \photoz\ and model-$z$ when assessing the agreement between them. We show them in Fig.~\ref{fig:modelz-photoz}.  

Ruling out \photoz\ solutions that place securely-identified lensed galaxies in front of the cluster, we find that the model-$z$ PDF of most of the sources are in good agreement with the \photoz\ PDF of at least one of the multiple images of that source. 
However, we note a discrepancy between the model-$z$ and the \photoz\ in the case of some of the sources and discuss in the following paragraph. 

The most problematic discrepancy is for source~9. The \hst\ colors and both the \hst\ and \hst+\spitzer\ \zphot\ PDFs rule out redshifts above 6, and the \photoz\ solutions of the two different images of the same source are in agreement. However, when the redshift of this system is set as a free parameter with a flat prior and no upper limit, all the lens models, including the ``bridge'' model, favor an extremely high redshift ($z\sim9$), albeit with large uncertainty. 
System~9 is a pair of images that closely straddle the critical curve. Such systems, if their spectroscopic redshift is known, can be excellent constraints, since they tightly constrain the location of the critical curve.
On the other hand, when the redshift of such a pair is unknown, only the position of the critical curve is constrained but not its redshift. Based on the colors and \photoz\ estimates for this source, we rule out the $z \sim 9$ solution. To examine the effect of this wrong solution on the lens model results, we computed a separate model with the redshift of system~9 fixed at $z=5.75$, 
the most probable \photoz\ of image 9.1 from the \hst+\spitzer\ \textsc{eazy} \photoz\ analysis.\footnote{The \hst\ \textsc{bpz} analysis yields \zphot\ $\sim 5.4$, thus this galaxy was not included as a high-z candidate in \cite{salmon17}.}
The outputs of the resulting model are not significantly different from models that leave this parameter free. Motivated by this examination, in our final model, we set the upper limit of the redshift of system~9 to $z\leq 6$.
}

The model predicted redshifts of sources 14 and 16 show a discrepancy with the \hst\ \textsc{bpz} PDF, however, the \hst+\spitzer\ \photoz\ increases the likelihood at higher redshifts, and their probability distributions do not rule out the model-$z$. Moreover, system 14 is faint (mag $\sim$ 28 - 29) and classified as \bronze, making this disagreement less concerning. 

For source 7, both \photoz\ analyses favor higher redshift solutions for this source, $z>3.5$, while the model-$z$ converges to $z\sim2.2$. The region in which this source appears is well constrained by images of sources~12 and 13. For 7.1 and 7.2 to be multiple images of the same source, source~7 must be at lower redshift than sources~12 and 13. If the \photoz\ is correct, this source may be misidentified, as already suggested by its classification as \bronze.

The \photoz\ PDFs of several systems, including systems 8 and 17 on the opposite side of systemas 9 and 14, indicate several solutions spanning a large range. Some of these solutions favor a higher redshift than predicted by the lens model. However, we cannot make definitive conclusions for such systems.  

Finally, we note that the photometric redshifts that were estimated from the \hst\ data alone were calculated using the \textsc{bpz} algorithm, and \hst+\spitzer\ photometric redshifts were calculated with \textsc{eazy}. While a thorough comparison of photometric redshifts is beyond the scope of this paper (e.g., \citealt{salmon17}), we show in Fig.~\ref{fig:modelz-photoz-EASY} in Appendix a similar comparison using the \textsc{eazy} algorithm for both the \hst\ and \hst+\spitzer\ photometric redshifts.

\begin{table*}[h!]
\centering
\caption{Best-Fit parameters for all three candidate mass models presented in this work.Are also given, their statistical estimators of the goodness of the best-fit, the rms, and the BIC. \label{tab:fit_param}}
\begin{tabular}{lrcccccccc}
\hline 
Model name & Component & $\Delta\alpha^{\rm ~a}$ & $\Delta\delta^{\rm ~a}$ & $\varepsilon^{\rm ~b}$ & $\theta^{\rm ~c}$ & $\sigma_0$ & r$_{\rm cut}$ & r$_{\rm core}$\\ 
(Fit statistics) & -- & (\arcsec) & (\arcsec) &   & ($\deg$) & (km\ s$^{-1}$) & (kpc) & (kpc)\\ 
\hline 
Silver constraints & DM & 6.2$^{+1.0}_{-0.8}$ & 9.1$^{+1.3}_{-0.9}$ & 0.78$^{+0.01}_{-0.01}$ & 54.2$^{+0.2}_{-0.3}$ & 1299.1$^{+16.9}_{-21.4}$ & [1500.0] & 32.8$^{+1.4}_{-1.2}$\\ 
rms = 0.37\arcsec\  & 1stBCG & [0.0] & [0.1] & [0.64] & [60.5] & 587.5$^{+2.7}_{-9.3}$ & 28.5$^{+13.2}_{-3.4}$ & 1.2$^{+0.2}_{-0.2}$\\ 
 BIC = 150 & 2ndBCG & [47.8] & [69.6] & [0.35] & [74.1] & 367.5$^{+14.7}_{-18.5}$ & 70.6$^{+18.2}_{-11.5}$ & 0.5$^{+0.4}_{-0.3}$\\ 
 & 3rdBCG & [46.9] & [48.4] & [0.16] & [50.6] & 256.5$^{+9.9}_{-13.9}$ & 74.9$^{+17.3}_{-24.1}$ & 0.2$^{+0.7}_{-0.1}$\\ 
  & $L^{*}$ Galaxy & -- & -- & -- & -- & 119.8$^{+9.7}_{-12.0}$ & -- & --\\ 
-- & -- & -- & -- & -- & -- & -- & -- & --\\ 
\hline 
\hline 
Bronze constraints & DM & 6.8$^{+0.4}_{-0.4}$ & 9.6$^{+0.6}_{-0.5}$ & 0.77$^{+0.02}_{-0.01}$ & 54.1$^{+0.3}_{-0.3}$ & 1284.2$^{+29.7}_{-35.5}$ & [1500.0] & 34.0$^{+0.4}_{-1.0}$\\ 
rms = 0.37\arcsec\  & 1stBCG & [0.0] & [0.1] & [0.64] & [60.5] & 597.0$^{+3.5}_{-8.7}$ & 41.0$^{+20.9}_{-16.3}$ & 1.6$^{+0.2}_{-0.2}$\\ 
 BIC = 164 & 2ndBCG & [47.8] & [69.6] & [0.35] & [74.1] & 394.8$^{+4.9}_{-15.1}$ & 55.4$^{+13.9}_{-13.9}$ & 1.0$^{+0.2}_{-0.3}$\\ 
 & 3rdBCG & [46.9] & [48.4] & [0.16] & [50.6] & 267.0$^{+17.2}_{-8.0}$ & 50.9$^{+21.6}_{-10.6}$ & 0.7$^{+0.7}_{-0.2}$\\ 
  & $L^{*}$ Galaxy & -- & -- & -- & -- & 116.0$^{+9.9}_{-17.1}$ & -- & --\\ 
 & -- & -- & -- & -- & -- & -- & -- & --\\ 
\hline 
\hline 
Bridge model & DM & 4.7$^{+2.2}_{-1.8}$ & 4.2$^{+4.4}_{-1.7}$ & 0.78$^{+0.02}_{-0.04}$ & 53.3$^{+0.4}_{-1.0}$ & 1037.0$^{+33.0}_{-136.3}$ & [1500.0] & 24.5$^{+2.0}_{-6.1}$\\ 
rms = 0.36\arcsec\  & bridge & 14.6$^{+1.8}_{-4.4}$ & 40.0$^{+0.1}_{-1.6}$ & 0.8$^{+0.25}_{-0.06}$ & 51.9$^{+37.9}_{-4.9}$ & 692.9$^{+126.7}_{-104.6}$ & [1500.0] & 45.9$^{+10.1}_{-3.8}$\\ 
 BIC = 172 & 1stBCG & [0.0] & [0.1] & [0.64] & [60.5] & 579.0$^{+14.2}_{-20.5}$ & 36.8$^{+11.7}_{-5.0}$ & 0.9$^{+0.3}_{-0.2}$\\ 
 & 2ndBCG & [47.8] & [69.6] & [0.35] & [74.1] & 379.7$^{+16.0}_{-8.8}$ & 62.1$^{+16.7}_{-8.8}$ & 0.5$^{+0.2}_{-0.1}$\\ 
   & 3rdBCG & [46.9] & [48.4] & [0.16] & [50.6] & 298.9$^{+12.7}_{-15.5}$ & 120.7$^{+2.5}_{-21.3}$ & 1.5$^{+0.0}_{-1.1}$\\ 
  & $L^{*}$ Galaxy & -- & -- & -- & -- & 94.4$^{+10.8}_{-11.9}$ & -- & --\\ 
-- & -- & -- & -- & -- & -- & -- & -- & --\\ 
\hline 

\end{tabular}
\medskip\\
$^{\rm a}$ $\Delta\alpha$ and $\Delta\delta$ are measured relative to the reference coordinate point: ($\alpha$ = 04:17:34.6925 , $\delta$ = -11:54:31.9356) \textbf{and given in arcsec going positively toward North West}.~~~~~~~~~~~~~~~~~~~~~~~~~~~~~~~~~~~~~~~~~~~~~~~~~~~~~~~~~~~~~~~~~~~~~~~~~~~~~~~~~~~~~~~~~~~~~~~~~~~\\[1pt]
$^{\rm b}$ Ellipticity ($\varepsilon$) is defined as $(a^2-b^2) / (a^2+b^2)$, where $a$ and $b$ are the semi-major and semi-minor axes respectively.~~~~~~~~~~~~~\\[1pt]
$^{\rm c}$ \textbf{$\theta$ goes counterclockwise from the West.}~~~~~~~~~~~~~~~~~~~~~~~~~~~~~~~~~~~~~~~~~~~~~~~~~~~~~~~~~~~~~~~~~~~~~~~~~~~~~~~~~~~~~~~~~~~~~~~~~~~~~~~~~\\[1pt]
Quantities in brackets are fixed parameters.~~~~~~~~~~~~~~~~~~~~~~~~~~~~~~~~~~~~~~~~~~~~~~~~~~~~~~~~~~~~~~~~~~~~~~~~~~~~~~~~~~~~~~~~~~~~~~~~~~~~~~~~~
\end{table*}

\section{High-Redshift Predictions}

During the first year of \JWST\ science operations,
at least 13 galaxy clusters will be observed in the context of the Guaranteed Time Observations (GTO) and {\it Director's Discretionary Early Release Science} (DD-ERS) programs (PIs: Windhorst, Willott, Stiavelli, Rigby, and Treu) using all four \JWST\ instruments: the Near-Infrared Camera (NIRCam), the Near-Infrared Imager and Slitless Spectrograph (NIRISS), Near-Infrared Spectrograph (NIRSpec), and the Mid-Infrared Instrument (MIRI).
These observations will include NIRCam imaging to various depths for all 13 clusters.
\clusterfullname\ is included in this list of 13 clusters and will be observed thanks to the Canadian NIRISS Unbiased Cluster Survey GTO program (CANUCS; PI: Willott). 

We use our lens model 
and UV luminosity functions from \citet{Mason2015}
to predict numbers of objects observable by \JWST\ 
at $8<z<16$, before and during the epoch of reionization.
We also explore and discuss the expectations from the \HST\ RELICS observations
that yielded 321 candidates with photometric redshifts \zphot\ $\sim 6 - 8$  in 46 cluster fields, but none from this cluster \citep{salmon17}.

\begin{figure}[h!]
\begin{center}
    \includegraphics[trim={0 0 6cm 0},clip,width=0.45\textwidth]{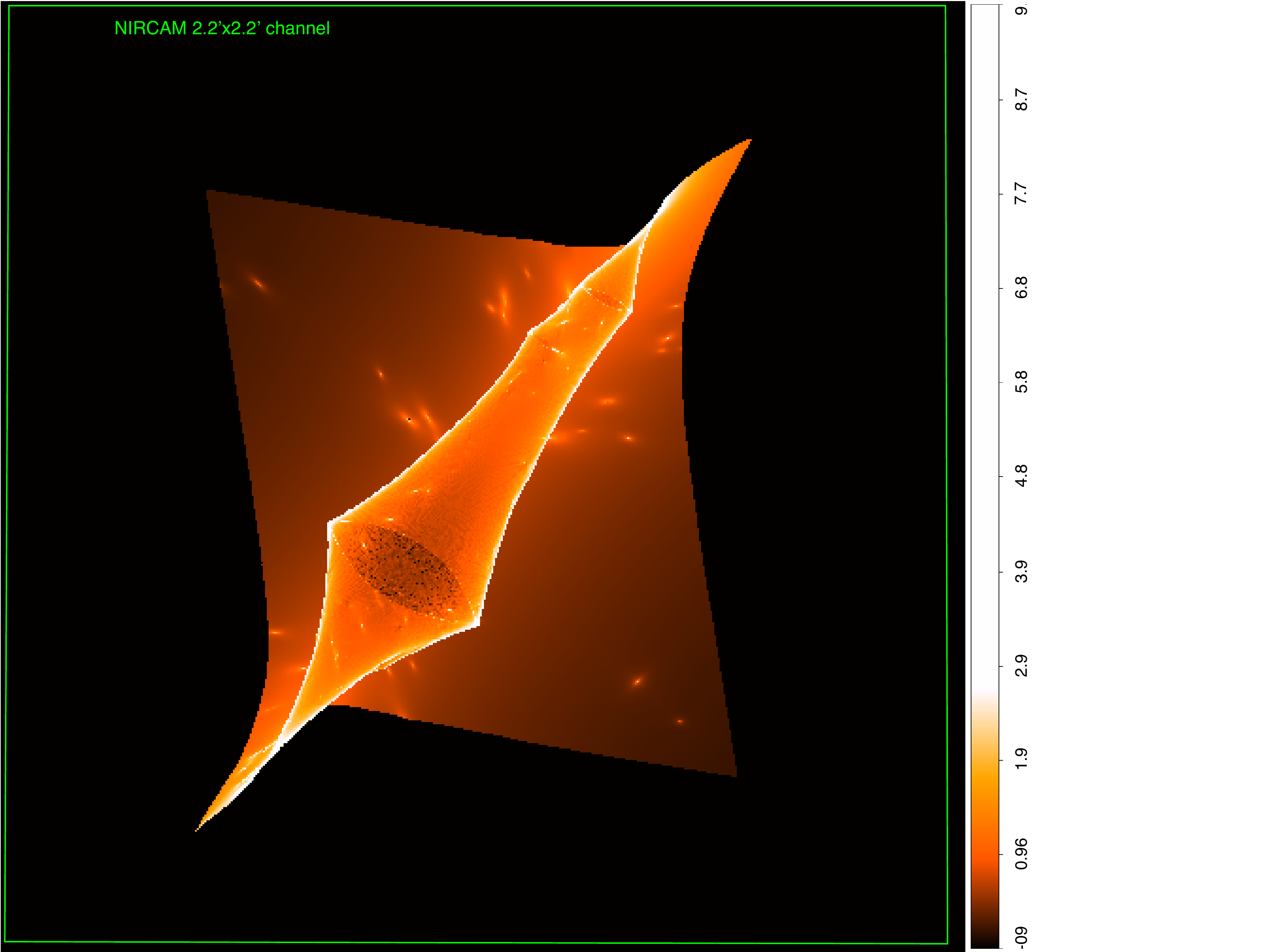}
        \caption{
        Delensed image of the \clustername\ magnification map for sources at $z=16$, 
        showing the source-plane area (1.3 arcmin$^2$) lensed into a $2\farcm2\times2\farcm2$ field of view (4.8 arcmin$^2$), field of view covered by a single NIRCam module. 
        The color scale shows the magnification factor in magnitudes.
        Beyond $z=7$, the delensed map does not differ significantly from the one presented here.
        }
    \label{fig:real_FoV}
    \end{center}
\end{figure}

Observing the high-redshift universe behind a cluster offers a boost in sensitivity to lower luminosities, but diminishes the field of view (FoV). 
In Fig.~\ref{fig:real_FoV}, we demonstrate how the effective observed FoV of $2\farcm2\times2\farcm2$ (4.8 arcmin$^2$, or one of the two modules observed by the \JWST/ NIRCam), is affected by gravitational lensing. The magnification map for a source at $z = 16$ is ray-traced through the best-fit model to the source plane. This transformation reveals the spatial extent of the background area covered by such an observation, resulting in an unlensed observed high-$z$ area of 1.3 arcmin$^2$.

\begin{figure}[h!]
\begin{center}
    \includegraphics[trim={0.5cm 6cm 0.5cm 6cm},clip,width=0.48\textwidth]{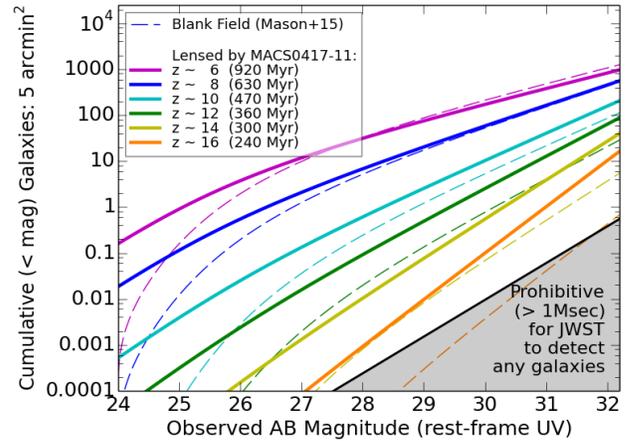}
        \caption{Cumulative number counts (not accounting for incompleteness) of galaxies expected at $z \sim 6$, 8, 10, 12, 14, and 16 in a 5\,arcmin$^2$ blank field (dashed lines) and lensed field (solid lines) based on luminosity functions from \citet{Mason2015} and our lens model of \clustername.   
        The black line very roughly assumes that a 1\,Msec program could detect galaxies with AB mag 32.2 in a single deep field and that the flux limit scales with $\sqrt{\rm exposure\ time}$ if that 1\,Msec is spread across a larger area.
        We expect strong lensing clusters such as these to deliver significant efficiency gains in discovering the first galaxies with \JWST,
        especially if luminosity function faint end slopes are as steep as predicted by \citet{Mason2015}.}
    \label{fig:count_obsmag}
    \end{center}
\end{figure}

Figure~\ref{fig:count_obsmag} shows the expected cumulative number counts (not accounting for incompleteness) for \clustername, or a galaxy cluster with similar lensing strength, as a function of magnitude,
for galaxies at $z=6$, $8$, $10$, $12$, $14$, and $16$
within the FoV of a single NIRCam module (roughly aligned with the WFC3/IR FoV). 
We adopt blank field luminosity functions from \citet{Mason2015} due to its ability to predict density at any redshifts.
The faint-end slope of this luminosity function increases from $\alpha = -2.1$ at $z = 8$ to $\alpha = -3.5$ at $z = 16$.
Such steep faint-end slopes would imply that many small, faint galaxies are magnified into view by lensing and that there is a significant efficiency gain from strong lensing to discover the first galaxies with \JWST.

Cluster observations scheduled for the first year of \JWST\ will typically reach magnitudes of $\sim$29\,AB and fainter.
From Fig.~\ref{fig:count_obsmag}, we expect that at this magnitude limit this field hosts three lensed galaxies at $z=10$, 
and less than one galaxy in each of the higher redshift bins, 
not accounting for detection efficiency and incompleteness. 
Observing of an order of a dozen clusters should yield galaxies as distant as $z=12$ and a substantial sample of high-$z$ galaxies at the epoch of reionization.


\begin{figure}[h!]
\begin{center}
    \includegraphics[trim={0.5cm 6cm 0.5cm 6cm},clip,width=0.45\textwidth]{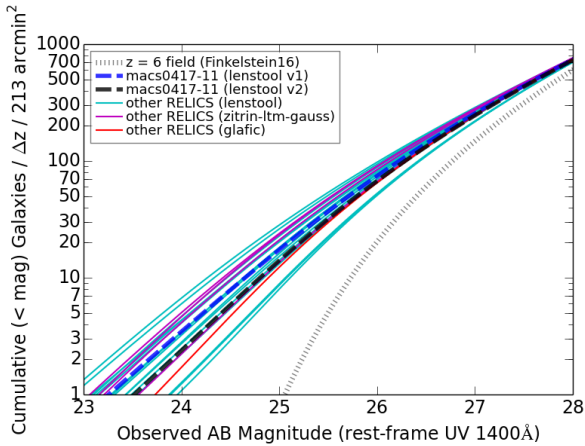}
        \caption{
        Expected number counts (not accounting for incompleteness) of $z = 6$ galaxies in blank fields (dashed line) or lensed by RELICS clusters according to our models (solid lines). 
        The first RELICS lens model (V1; dark blue line) predicts MACS\,J0417 to have a relatively average lensing strength compared to other RELICS clusters. On the other hand, the model presented in this paper (V2; black line) is among the 25\% weakest 
        of 21 RELICS clusters for which mass models are already available.
        All expectations are scaled to the full area of 213\,arcmin$^2$ covered on the sky by RELICS.
        The publicly available lens models were derived with various methods:
        \lenstool\ \citep{Kneib1996,Jullo2007}, 
        Zitrin-LTM \citep{Broadhurst2005,Zitrin2015},
        and GLAFIC \citep{Oguri2010}.
        }
    \label{fig:full-RELICS}
    \end{center}
\end{figure}

In Fig.~\ref{fig:full-RELICS}, we compare the lensing strength of \clustername\ to other clusters from the RELICS program for which lens models are publicly available on the MAST, including those published by \citet{cerny18,Acebron2018a,Acebron2018b,Cibirka2018,Paterno-Mahler2018}.
The previous version of \clustername\ lens model, V1, available on the MAST, predicts $\sim 20 \%$ higher number counts for relatively bright sources (AB mag 25), and similar number counts for faint sources, giving an indication of the systematic uncertainties due to spectroscopic redshift availability, and different modeling assumptions.

With the updated model (V2), we find that \clustername\ is ranked in the lower 25th percentile of these clusters when it comes to the lensing strength. 
However, as other RELICS clusters, MACS0417 is among the most powerful lenses known to date.

In a photometric search for $z \sim 6 - 8$ galaxies in the entire RELICS survey, \citet{salmon17} report 321 candidates, with a median of six candidates per field and an average of seven, none of which are in the field of \clustername. 
From Poisson statistics alone, there is a 4\% chance that at least one of the 46 RELICS fields would yield no $z \sim 6 - 8$ candidates, given the average of seven per field. Cosmic variance would increase this likelihood somewhat \citep{Trenti08}, especially in a lensed field \citep{Robertson14}. 
However, our lensing analysis indicates that the lensing strength of \clustername\ is not extraordinarily low compared to other RELICS clusters for which models are available. It is therefore odd that \citet{salmon17} detected no \zphot $\sim 6 - 8$ candidates in this field. 

Quantitatively, the prediction for \clustername, shown in Fig.~\ref{fig:full-RELICS},
indicates that this field should host  
about 5.34 $z \sim 6$ magnified galaxies at, or brighter than, 27 mag.
The actual expected number would be lower, due to incompleteness.
A thorough investigation, including completeness estimates, is required (e.g., \citealt{Livermore2017}). 
However, we can get a rough estimate of the detection efficiency of \citet{salmon17} for the discovery of 
$z\sim6$ galaxies from their actual detection histograms. \citet{salmon17} discovered 211 candidates with F160W AB mag $\leq 27$ in the \zphot $=6$ bin in all of the RELICS fields. From Fig.~\ref{fig:full-RELICS}, we expect there to be at most 300 galaxies at $z=6$ with observed AB magnitude below 27 within the same observed area. A comparison of the number of candidates observed with the predicted number implies an estimated {\it average} efficiency of at least $70\%$.
Assuming this efficiency, we would have expected \citet{salmon17} to find at least 5.34$\times 70\% = 3.74 $ galaxies in this range behind \clustername. 
Assuming small-number statistics, the zero detection is discrepant with this estimated expectation (for example, Poisson statistics would give a range of 1--8 at 95\% confidence level).  
The low number of candidates in this field could be a result of lower-than-average density of galaxies at this location due to cosmic variance. 
However, such discrepancy suggests a reanalysis of this particular field is needed.  

As can be seen in Fig.~\ref{fig:modelz-photoz} and \ref{fig:modelz-photoz-EASY}, some of the \textsc{eazy} \photoz\ PDFs favor $z>5.5$ solutions for some of the multiple images. A preliminary \textsc{bpz} reanalysis of this field puts source~9 slightly above \zphot$=5.5$, which would increase the number of candidates in this field to two $z \sim 6$ candidates. 
Therefore reducing the disagreement between predictions and detections.

An analysis of this field and all RELICS fields based on the combined \hst+\spitzer\ photometry is in progress (Strait et al.\ in prep.). Adding the \spitzer\ photometry could remove some of the degeneracies and improve the photometric redshift estimates.


\section{Discussion and summary}
\label{sec:discussion}

We present a strong lens model of \clusterfullname, updating the model previously released by the RELICS collaboration. This cluster was selected by the RELICS program for its promising lensing capabilities. We identified 57 multiple images belonging to 17 lensed background sources. 
We also report lensing candidates that were not reliable enough to be used as constraints but are nevertheless of potential interest for further study by current or upcoming facilities such as \JWST. 
{This study and the companion paper by \cite{Jauzac2018c} represent the first published strong lensing analyses of this cluster.}

{Our strong lensing analysis compares models based on constraints with different levels of reliability (\silver\ and \bronze) as well as different levels of complexity of the lens plane, for example when including a bridge of matter between the two main substructures of the cluster. Our analysis reveals that the addition of a {\it bridge} potential, while giving a lower rms does not satisfy our BIC criteria.  Therefore we keep a fiducial model constrained by our \silver\ sample with no potential acting as a bridge of matter between substructures of the cluster.}

From our strong lensing mass modeling, we measure a total projected mass within 200\,kpc of \Mtwokpc.
Using the parametric capability of our modeling we estimate the mass ratio between the large scale halo and the galaxy halos to be of order 100:1. 
Extrapolating the mass model to a large projected radius, we find a mass at 1.5\,Mpc of \MfiveMpc. 
Despite the limited ability of strong lens models to measure the mass beyond the multiple image region, this value is within 3$\sigma$ of \textbf{the mass $M_{1.5Mpc} = (18.9\pm0.25) \times10^{14} M_\odot$ measured by weak lensing analysis \citep{Applegate2014}.
We report for this cluster an Einstein radius of \Erad\ at $z=9$. 
Using the parameters of the spherical Navarro--Frenk--White (NFW,\citealt{NFW}) profile fitted in \cite{Applegate2014}, we derived an Einstein radius at $z=9$ of $\theta_{E_{NFW}}\simeq$26\arcsec . The large mass reported \cite{Applegate2014} still provide a reasonably close Einstein radius compare to our analysis. In addition, strong lensing analysis of CLASH clusters \citep{Zitrin2015} report for comparable clusters similar values.
}

{We examine the agreement between \photoz\ and model-$z$ for the sample of multiple images selected in our study. There is a general agreement when the low-$z$ solutions for the \photoz\ are excluded. System 7 might be a mis-identification. The agreements for systems 12 and 13 benefit from the reduced redshift range during the optimization of system 9 induced by the initial disagreement with \photoz\ . A detailed study of the influence of the photometric redshift algorithm or the dataset is beyond the scope of this paper as this would need more spectroscopic redshifts to be used as a benchmark to remove biases in the comparison.}

{Our previous model of \clustername\ suggested its lensing strength was about average among all RELICS clusters modeled to date (all of which are powerful lenses).
The new lens model presented here suggests \clustername\ is in the lower 25th percentile of the RELICS clusters.
Still the lack of any \zphot\ $\sim 6 - 8$ candidates in this field is at odds with the expected number, estimated from the lensing magnification of this field, assumptions on the high-$z$ luminosity functions, and our estimate of the average detection efficiency of \citet{salmon17}.
We primarily attribute this to cosmic variance, but we will reanalyze this field and perform completeness simulations to determine if there are some other reasons besides cosmic variance for the low yield of high-z candidates.
\clustername\ is still expected to be an efficient lens for upcoming \JWST\ GTO observations to discover fainter and higher redshift candidates.
Strong lensing clusters will continue to deliver significant efficiency gains toward discovering high-redshift galaxies and the first galaxies with \JWST.}

\section*{Acknowledgments}

Support for program GO-14096 was provided by NASA through a grant from the Space Telescope Science Institute, which is operated by the Association of Universities for Research in Astronomy, Inc., under NASA contract NAS5-26555. This paper is based on observations made with the NASA/ESA Hubble Space Telescope, obtained at the Space Telescope Science Institute, which is operated by the Association of Universities for Research in Astronomy, Inc., under NASA contract NAS 5-26555. These observations are associated with program GO-14096. Archival data are associated with program GO-12009. 
This paper includes data gathered with the 6.5 meter Magellan Telescopes located at Las Campanas Observatory, Chile.
MJ was supported by the Science and Technology Facilities Council (grant number ST/L00075X/1) and used the DiRAC Data Centric system at Durham University, operated by the Institute for Computational Cosmology on behalf of the STFC DiRAC HPC Facility (\url{www.dirac.ac.uk}). This equipment was funded by BIS National E-infrastructure capital grant ST/K00042X/1, STFC capital grant ST/H008519/1, and STFC DiRAC Operations grant ST/K003267/1 and Durham University. DiRAC is part of the National E-Infrastructure.
ACE acknowledges support from STFC grant ST/P00541/1.
IUR acknowledges support from NSF grants AST-1613536 and
AST-1815403.
Lawrence Livermore National Laboratory is operated by Lawrence Livermore National Security, LLC, for the U.S. Department of Energy, National Nuclear Security Administration under Contract DE-AC52-07NA27344
%

\facilities{HST(WFC3,ACS) Magellan(LDSS3), VLT(MUSE)}



\bibliography{biblio_MACS0417}

\begin{thebibliography}{}
\expandafter\ifx\csname natexlab\endcsname\relax\def\natexlab#1{#1}\fi
\providecommand{\url}[1]{\href{#1}{#1}}

\bibitem[{{Acebron} {et~al.}(2017){Acebron}, {Jullo}, {Limousin}, {Tilquin},
  {Giocoli}, {Jauzac}, {Mahler}, \& {Richard}}]{Acebron2017}
{Acebron}, A., {Jullo}, E., {Limousin}, M., {et~al.} 2017, \mnras, 470, 1809

\bibitem[{{Acebron} {et~al.}(2018{\natexlab{a}}){Acebron}, {Cibirka}, {Zitrin},
  {Coe}, {Agulli}, {Sharon}, {Brada{\v{c}}}, {Frye}, {Livermore}, {Mahler},
  {Salmon}, {Umetsu}, {Bradley}, {Andrade-Santos}, {Avila}, {Carrasco},
  {Cerny}, {Czakon}, {Dawson}, {Hoag}, {Huang}, {Johnson}, {Jones},
  {Kikuchihara}, {Lam}, {Lovisari}, {Mainali}, {Oesch}, {Ogaz}, {Ouchi},
  {Past}, {Paterno-Mahler}, {Peterson}, {Ryan}, {Sendra-Server}, {Stark},
  {Strait}, {Toft}, {Trenti}, \& {Vulcani}}]{Acebron2018a}
{Acebron}, A., {Cibirka}, N., {Zitrin}, A., {et~al.} 2018{\natexlab{a}}, \apj,
  858, 42

\bibitem[{{Acebron} {et~al.}(2018{\natexlab{b}}){Acebron}, {Alon}, {Zitrin},
  {Mahler}, {Coe}, {Sharon}, {Cibirka}, {Brada{\v{c}}}, {Trenti}, {Umetsu},
  {Andrade-Santos}, {Avila}, {Bradley}, {Carrasco}, {Cerny}, {Czakon},
  {Dawson}, {Frye}, {Hoag}, {Huang}, {Johnson}, {Jones}, {Kikuchihara}, {Lam},
  {Livermore}, {Lovisari}, {Mainali}, {Oesch}, {Ogaz}, {Ouchi}, {Past},
  {Paterno-Mahler}, {Peterson}, {Ryan}, {Salmon}, {Sendra- Server}, {Stark},
  {Strait}, {Toft}, \& {Vulcani}}]{Acebron2018b}
{Acebron}, A., {Alon}, M., {Zitrin}, A., {et~al.} 2018{\natexlab{b}}, arXiv
  e-prints, arXiv:1810.08122

\bibitem[{{Applegate} {et~al.}(2014){Applegate}, {von der Linden}, {Kelly},
  {Allen}, {Allen}, {Burchat}, {Burke}, {Ebeling}, {Mantz}, \&
  {Morris}}]{Applegate2014}
{Applegate}, D.~E., {von der Linden}, A., {Kelly}, P.~L., {et~al.} 2014,
  \mnras, 439, 48

\bibitem[{{Atek} {et~al.}(2018){Atek}, {Richard}, {Kneib}, \&
  {Schaerer}}]{Atek2018}
{Atek}, H., {Richard}, J., {Kneib}, J.-P., \& {Schaerer}, D. 2018, \mnras, 479,
  5184

\bibitem[{{Atek} {et~al.}(2015){Atek}, {Richard}, {Kneib}, {Jauzac},
  {Schaerer}, {Clement}, {Limousin}, {Jullo}, {Natarajan}, {Egami}, \&
  {Ebeling}}]{Atek2015}
{Atek}, H., {Richard}, J., {Kneib}, J.-P., {et~al.} 2015, \apj, 800, 18

\bibitem[{{Bacon} {et~al.}(2010){Bacon}, {Accardo}, {Adjali}, {Anwand},
  {Bauer}, {Biswas}, {Blaizot}, {Boudon}, {Brau-Nogue}, {Brinchmann},
  {Caillier}, {Capoani}, {Carollo}, {Contini}, {Couderc}, {Daguis{\'e}},
  {Deiries}, {Delabre}, {Dreizler}, {Dubois}, {Dupieux}, {Dupuy}, {Emsellem},
  {Fechner}, {Fleischmann}, {Fran{\c c}ois}, {Gallou}, {Gharsa}, {Glindemann},
  {Gojak}, {Guiderdoni}, {Hansali}, {Hahn}, {Jarno}, {Kelz}, {Koehler},
  {Kosmalski}, {Laurent}, {Le Floch}, {Lilly}, {Lizon}, {Loupias}, {Manescau},
  {Monstein}, {Nicklas}, {Olaya}, {Pares}, {Pasquini}, {P{\'e}contal-Rousset},
  {Pell{\'o}}, {Petit}, {Popow}, {Reiss}, {Remillieux}, {Renault}, {Roth},
  {Rupprecht}, {Serre}, {Schaye}, {Soucail}, {Steinmetz}, {Streicher}, {Stuik},
  {Valentin}, {Vernet}, {Weilbacher}, {Wisotzki}, \& {Yerle}}]{Bacon2010}
{Bacon}, R., {Accardo}, M., {Adjali}, L., {et~al.} 2010, in \procspie, Vol.
  7735, Ground-based and Airborne Instrumentation for Astronomy III, 773508

\bibitem[{{Becker} {et~al.}(2015){Becker}, {Bolton}, \& {Lidz}}]{Becker2015}
{Becker}, G.~D., {Bolton}, J.~S., \& {Lidz}, A. 2015, \pasa, 32, e045

\bibitem[{{Becker} {et~al.}(2001){Becker}, {Fan}, {White}, {Strauss},
  {Narayanan}, {Lupton}, {Gunn}, {Annis}, {Bahcall}, {Brinkmann}, {Connolly},
  {Csabai}, {Czarapata}, {Doi}, {Heckman}, {Hennessy}, {Ivezi{\'c}}, {Knapp},
  {Lamb}, {McKay}, {Munn}, {Nash}, {Nichol}, {Pier}, {Richards}, {Schneider},
  {Stoughton}, {Szalay}, {Thakar}, \& {York}}]{Becker2001}
{Becker}, R.~H., {Fan}, X., {White}, R.~L., {et~al.} 2001, \aj, 122, 2850

\bibitem[{{Ben{\'{\i}}tez}(2000)}]{Benitez00}
{Ben{\'{\i}}tez}, N. 2000, \apj, 536, 571

\bibitem[{{Bertin} \& {Arnouts}(1996)}]{SEx}
{Bertin}, E., \& {Arnouts}, S. 1996, \aaps, 117, 393

\bibitem[{{Bhatawdekar} {et~al.}(2018){Bhatawdekar}, {Conselice}, {Margalef-
  Bentabol}, \& {Duncan}}]{Bhatawdekar2018}
{Bhatawdekar}, R., {Conselice}, C., {Margalef- Bentabol}, B., \& {Duncan}, K.
  2018, ArXiv e-prints, arXiv:1807.07580

\bibitem[{{Botteon} {et~al.}(2018){Botteon}, {Gastaldello}, \&
  {Brunetti}}]{Botteon2018}
{Botteon}, A., {Gastaldello}, F., \& {Brunetti}, G. 2018, \mnras, 476, 5591

\bibitem[{{Bouwens} {et~al.}(2017){Bouwens}, {Oesch}, {Illingworth}, {Ellis},
  \& {Stefanon}}]{Bouwens2017}
{Bouwens}, R.~J., {Oesch}, P.~A., {Illingworth}, G.~D., {Ellis}, R.~S., \&
  {Stefanon}, M. 2017, \apj, 843, 129

\bibitem[{{Bradley} {et~al.}(2014){Bradley}, {Zitrin}, {Coe}, {Bouwens},
  {Postman}, {Balestra}, {Grillo}, {Monna}, {Rosati}, {Seitz}, {Host}, {Lemze},
  {Moustakas}, {Moustakas}, {Shu}, {Zheng}, {Broadhurst}, {Carrasco}, {Jouvel},
  {Koekemoer}, {Medezinski}, {Meneghetti}, {Nonino}, {Smit}, {Umetsu},
  {Bartelmann}, {Ben{\'{\i}}tez}, {Donahue}, {Ford}, {Infante}, {Jimenez-Teja},
  {Kelson}, {Lahav}, {Maoz}, {Melchior}, {Merten}, \& {Molino}}]{Bradley2014}
{Bradley}, L.~D., {Zitrin}, A., {Coe}, D., {et~al.} 2014, \apj, 792, 76

\bibitem[{{Brammer} {et~al.}(2008){Brammer}, {van Dokkum}, \&
  {Coppi}}]{Brammer2008}
{Brammer}, G.~B., {van Dokkum}, P.~G., \& {Coppi}, P. 2008, \apj, 686, 1503

\bibitem[{{Broadhurst} {et~al.}(2005){Broadhurst}, {Ben{\'{\i}}tez}, {Coe},
  {Sharon}, {Zekser}, {White}, {Ford}, {Bouwens}, {Blakeslee}, {Clampin},
  {Cross}, {Franx}, {Frye}, {Hartig}, {Illingworth}, {Infante}, {Menanteau},
  {Meurer}, {Postman}, {Ardila}, {Bartko}, {Brown}, {Burrows}, {Cheng},
  {Feldman}, {Golimowski}, {Goto}, {Gronwall}, {Herranz}, {Holden}, {Homeier},
  {Krist}, {Lesser}, {Martel}, {Miley}, {Rosati}, {Sirianni}, {Sparks},
  {Steindling}, {Tran}, {Tsvetanov}, \& {Zheng}}]{Broadhurst2005}
{Broadhurst}, T., {Ben{\'{\i}}tez}, N., {Coe}, D., {et~al.} 2005, \apj, 621, 53

\bibitem[{{Cerny} {et~al.}(2018){Cerny}, {Sharon}, {Andrade-Santos}, {Avila},
  {Brada{\v c}}, {Bradley}, {Carrasco}, {Coe}, {Czakon}, {Dawson}, {Frye},
  {Hoag}, {Huang}, {Johnson}, {Jones}, {Lam}, {Lovisari}, {Mainali}, {Oesch},
  {Ogaz}, {Past}, {Paterno-Mahler}, {Peterson}, {Riess}, {Rodney}, {Ryan},
  {Salmon}, {Sendra-Server}, {Stark}, {Strolger}, {Trenti}, {Umetsu},
  {Vulcani}, \& {Zitrin}}]{cerny18}
{Cerny}, C., {Sharon}, K., {Andrade-Santos}, F., {et~al.} 2018, \apj, 859, 159

\bibitem[{{Chiriv{\`i}} {et~al.}(2018){Chiriv{\`i}}, {Suyu}, {Grillo},
  {Halkola}, {Balestra}, {Caminha}, {Mercurio}, \& {Rosati}}]{Chirivi2018}
{Chiriv{\`i}}, G., {Suyu}, S.~H., {Grillo}, C., {et~al.} 2018, \aap, 614, A8

\bibitem[{{Cibirka} {et~al.}(2018){Cibirka}, {Acebron}, {Zitrin}, {Coe},
  {Agulli}, {Andrade-Santos}, {Brada{\v c}}, {Frye}, {Livermore}, {Mahler},
  {Salmon}, {Sharon}, {Trenti}, {Umetsu}, {Avila}, {Bradley}, {Carrasco},
  {Cerny}, {Czakon}, {Dawson}, {Hoag}, {Huang}, {Johnson}, {Jones},
  {Kikuchihara}, {Lam}, {Lovisari}, {Mainali}, {Oesch}, {Ogaz}, {Ouchi},
  {Past}, {Paterno-Mahler}, {Peterson}, {Ryan}, {Sendra-Server}, {Stark},
  {Strait}, {Toft}, \& {Vulcani}}]{Cibirka2018}
{Cibirka}, N., {Acebron}, A., {Zitrin}, A., {et~al.} 2018, \apj, 863, 145

\bibitem[{{Coe} {et~al.}(2006){Coe}, {Ben{\'{\i}}tez}, {S{\'a}nchez}, {Jee},
  {Bouwens}, \& {Ford}}]{Coe06}
{Coe}, D., {Ben{\'{\i}}tez}, N., {S{\'a}nchez}, S.~F., {et~al.} 2006, \aj, 132,
  926

\bibitem[{{Coe} {et~al.}(2013){Coe}, {Zitrin}, {Carrasco}, {Shu}, {Zheng},
  {Postman}, {Bradley}, {Koekemoer}, {Bouwens}, {Broadhurst}, {Monna}, {Host},
  {Moustakas}, {Ford}, {Moustakas}, {van der Wel}, {Donahue}, {Rodney},
  {Ben{\'{\i}}tez}, {Jouvel}, {Seitz}, {Kelson}, \& {Rosati}}]{Coe2013}
{Coe}, D., {Zitrin}, A., {Carrasco}, M., {et~al.} 2013, \apj, 762, 32

\bibitem[{{Djorgovski} {et~al.}(2001){Djorgovski}, {Castro}, {Stern}, \&
  {Mahabal}}]{Djorgovski2001}
{Djorgovski}, S.~G., {Castro}, S., {Stern}, D., \& {Mahabal}, A.~A. 2001, \apj,
  560, L5

\bibitem[{{Dressler} {et~al.}(2011){Dressler}, {Bigelow}, {Hare}, {Sutin},
  {Thompson}, {Burley}, {Epps}, {Oemler}, {Bagish}, {Birk}, {Clardy},
  {Gunnels}, {Kelson}, {Shectman}, \& {Osip}}]{Dressler2011}
{Dressler}, A., {Bigelow}, B., {Hare}, T., {et~al.} 2011, \pasp, 123, 288

\bibitem[{{Dwarakanath} {et~al.}(2011){Dwarakanath}, {Malu}, \&
  {Kale}}]{Dwarakanath2011}
{Dwarakanath}, K.~S., {Malu}, S., \& {Kale}, R. 2011, Journal of Astrophysics
  and Astronomy, 32, 529

\bibitem[{{Ebeling} {et~al.}(2001){Ebeling}, {Edge}, \& {Henry}}]{Ebeling2001}
{Ebeling}, H., {Edge}, A.~C., \& {Henry}, J.~P. 2001, \apj, 553, 668

\bibitem[{{Ebeling} {et~al.}(2014){Ebeling}, {Stephenson}, \&
  {Edge}}]{Ebeling2014}
{Ebeling}, H., {Stephenson}, L.~N., \& {Edge}, A.~C. 2014, \apjl, 781, L40

\bibitem[{{El{\'{\i}}asd{\'o}ttir} {et~al.}(2007){El{\'{\i}}asd{\'o}ttir},
  {Limousin}, {Richard}, {Hjorth}, {Kneib}, {Natarajan}, {Pedersen}, {Jullo},
  \& {Paraficz}}]{Eliasdottir2007}
{El{\'{\i}}asd{\'o}ttir}, {\'A}., {Limousin}, M., {Richard}, J., {et~al.} 2007,
  ArXiv e-prints, arXiv:0710.5636

\bibitem[{{Fan} {et~al.}(2006){Fan}, {Strauss}, {Becker}, {White}, {Gunn},
  {Knapp}, {Richards}, {Schneider}, {Brinkmann}, \& {Fukugita}}]{Fan2006}
{Fan}, X., {Strauss}, M.~A., {Becker}, R.~H., {et~al.} 2006, \aj, 132, 117

\bibitem[{{Girardi} {et~al.}(2002){Girardi}, {Manzato}, {Mezzetti}, {Giuricin},
  \& {Limboz}}]{Girardi2002}
{Girardi}, M., {Manzato}, P., {Mezzetti}, M., {Giuricin}, G., \& {Limboz}, F.
  2002, \apj, 569, 720

\bibitem[{{Gladders} \& {Yee}(2000)}]{Gladders2000}
{Gladders}, M.~D., \& {Yee}, H.~K.~C. 2000, \aj, 120, 2148

\bibitem[{{Green} {et~al.}(2016){Green}, {Edge}, {Stott}, {Ebeling}, {Burgett},
  {Chambers}, {Draper}, {Metcalfe}, {Kaiser}, {Wainscoat}, \&
  {Waters}}]{Green2016}
{Green}, T.~S., {Edge}, A.~C., {Stott}, J.~P., {et~al.} 2016, \mnras, 461, 560

\bibitem[{{Gunn} \& {Peterson}(1965)}]{Gunn1965}
{Gunn}, J.~E., \& {Peterson}, B.~A. 1965, \apj, 142, 1633

\bibitem[{{Hashimoto} {et~al.}(2018){Hashimoto}, {Laporte}, {Mawatari},
  {Ellis}, {Inoue}, {Zackrisson}, {Roberts-Borsani}, {Zheng}, {Tamura},
  {Bauer}, {Fletcher}, {Harikane}, {Hatsukade}, {Hayatsu}, {Matsuda}, {Matsuo},
  {Okamoto}, {Ouchi}, {Pell{\'o}}, {Rydberg}, {Shimizu}, {Taniguchi},
  {Umehata}, \& {Yoshida}}]{Hashimoto2018}
{Hashimoto}, T., {Laporte}, N., {Mawatari}, K., {et~al.} 2018, \nat, 557, 392

\bibitem[{{Ishigaki} {et~al.}(2018){Ishigaki}, {Kawamata}, {Ouchi}, {Oguri},
  {Shimasaku}, \& {Ono}}]{Ishigaki2018}
{Ishigaki}, M., {Kawamata}, R., {Ouchi}, M., {et~al.} 2018, \apj, 854, 73

\bibitem[{{Jauzac} {et~al.}(2018){Jauzac}, {Harvey}, \& {Massey}}]{Jauzac2018a}
{Jauzac}, M., {Harvey}, D., \& {Massey}, R. 2018, \mnras, 477, 4046

\bibitem[{{Jauzac} {et~al.}(2015){Jauzac}, {Richard}, {Jullo}, {Cl{\'e}ment},
  {Limousin}, {Kneib}, {Ebeling}, {Natarajan}, {Rodney}, {Atek}, {Massey},
  {Eckert}, {Egami}, \& {Rexroth}}]{Jauzac2015}
{Jauzac}, M., {Richard}, J., {Jullo}, E., {et~al.} 2015, \mnras, 452, 1437

\bibitem[{{Jauzac} {et~al.}(2019){Jauzac}, {Mahler}, {Edge}, {Sharon},
  {Gillman}, {Ebeling}, {Harvey}, {Richard}, {Hamer}, {Fumagalli}, {Mark
  Swinbank}, {Kneib}, {Massey}, \& {Salom{\'e}}}]{Jauzac2018c}
{Jauzac}, M., {Mahler}, G., {Edge}, A.~C., {et~al.} 2019, \mnras, 483, 3082

\bibitem[{{Johnson} \& {Sharon}(2016)}]{Johnson2016}
{Johnson}, T.~L., \& {Sharon}, K. 2016, ArXiv e-prints, arXiv:1608.08713

\bibitem[{{Jullo} {et~al.}(2007){Jullo}, {Kneib}, {Limousin},
  {El{\'{\i}}asd{\'o}ttir}, {Marshall}, \& {Verdugo}}]{Jullo2007}
{Jullo}, E., {Kneib}, J.-P., {Limousin}, M., {et~al.} 2007, New Journal of
  Physics, 9, 447

\bibitem[{{Kneib} {et~al.}(1996){Kneib}, {Ellis}, {Smail}, {Couch}, \&
  {Sharples}}]{Kneib1996}
{Kneib}, J.-P., {Ellis}, R.~S., {Smail}, I., {Couch}, W.~J., \& {Sharples},
  R.~M. 1996, \apj, 471, 643

\bibitem[{{Lagattuta} {et~al.}(2017){Lagattuta}, {Richard}, {Cl{\'e}ment},
  {Mahler}, {Patr{\'{\i}}cio}, {Pell{\'o}}, {Soucail}, {Schmidt}, {Wisotzki},
  {Martinez}, \& {Bina}}]{Lagattuta2017}
{Lagattuta}, D.~J., {Richard}, J., {Cl{\'e}ment}, B., {et~al.} 2017, \mnras,
  469, 3946

\bibitem[{{Limousin} {et~al.}(2005){Limousin}, {Kneib}, \&
  {Natarajan}}]{Limousin2005}
{Limousin}, M., {Kneib}, J.-P., \& {Natarajan}, P. 2005, \mnras, 356, 309

\bibitem[{{Limousin} {et~al.}(2010){Limousin}, {Jullo}, {Richard}, {Cabanac},
  {Suyu}, {Halkola}, {Kneib}, {Gavazzi}, \& {Soucail}}]{Limousin2010}
{Limousin}, M., {Jullo}, E., {Richard}, J., {et~al.} 2010, \aap, 524, A95

\bibitem[{{Livermore} {et~al.}(2016){Livermore}, {Finkelstein}, \&
  {Lotz}}]{Livermore2016}
{Livermore}, R.~C., {Finkelstein}, S.~L., \& {Lotz}, J.~M. 2016, ArXiv
  e-prints, arXiv:1604.06799

\bibitem[{{Livermore} {et~al.}(2017){Livermore}, {Finkelstein}, \&
  {Lotz}}]{Livermore2017}
---. 2017, \apj, 835, 113

\bibitem[{{Lotz} {et~al.}(2017){Lotz}, {Koekemoer}, {Coe}, {Grogin}, {Capak},
  {Mack}, {Anderson}, {Avila}, {Barker}, {Borncamp}, {Brammer}, {Durbin},
  {Gunning}, {Hilbert}, {Jenkner}, {Khandrika}, {Levay}, {Lucas}, {MacKenty},
  {Ogaz}, {Porterfield}, {Reid}, {Robberto}, {Royle}, {Smith},
  {Storrie-Lombardi}, {Sunnquist}, {Surace}, {Taylor}, {Williams}, {Bullock},
  {Dickinson}, {Finkelstein}, {Natarajan}, {Richard}, {Robertson}, {Tumlinson},
  {Zitrin}, {Flanagan}, {Sembach}, {Soifer}, \& {Mountain}}]{Lotz2017}
{Lotz}, J.~M., {Koekemoer}, A., {Coe}, D., {et~al.} 2017, \apj, 837, 97

\bibitem[{{Mahler} {et~al.}(2018){Mahler}, {Richard}, {Cl{\'e}ment},
  {Lagattuta}, {Schmidt}, {Patr{\'{\i}}cio}, {Soucail}, {Bacon}, {Pello},
  {Bouwens}, {Maseda}, {Martinez}, {Carollo}, {Inami}, {Leclercq}, \&
  {Wisotzki}}]{Mahler2018}
{Mahler}, G., {Richard}, J., {Cl{\'e}ment}, B., {et~al.} 2018, \mnras, 473, 663

\bibitem[{{Mann} \& {Ebeling}(2012)}]{Mann2012}
{Mann}, A.~W., \& {Ebeling}, H. 2012, \mnras, 420, 2120

\bibitem[{{Mason} {et~al.}(2015){Mason}, {Trenti}, \& {Treu}}]{Mason2015}
{Mason}, C.~A., {Trenti}, M., \& {Treu}, T. 2015, \apj, 813, 21

\bibitem[{{Meneghetti} {et~al.}(2017){Meneghetti}, {Natarajan}, {Coe},
  {Contini}, {De Lucia}, {Giocoli}, {Acebron}, {Borgani}, {Bradac}, {Diego},
  {Hoag}, {Ishigaki}, {Johnson}, {Jullo}, {Kawamata}, {Lam}, {Limousin},
  {Liesenborgs}, {Oguri}, {Sebesta}, {Sharon}, {Williams}, \&
  {Zitrin}}]{Meneghetti2017}
{Meneghetti}, M., {Natarajan}, P., {Coe}, D., {et~al.} 2017, \mnras, 472, 3177

\bibitem[{{Merlin} {et~al.}(2015){Merlin}, {Fontana}, {Ferguson}, {Dunlop},
  {Elbaz}, {Bourne}, {Bruce}, {Buitrago}, {Castellano}, {Schreiber}, {Grazian},
  {McLure}, {Okumura}, {Shu}, {Wang}, {Amor{\'\i}n}, {Boutsia}, {Cappelluti},
  {Comastri}, {Derriere}, {Faber}, \& {Santini}}]{merl15}
{Merlin}, E., {Fontana}, A., {Ferguson}, H.~C., {et~al.} 2015, \aap, 582, A15

\bibitem[{{Merlin} {et~al.}(2016){Merlin}, {Fontana}, {Ferguson}, {Dunlop},
  {Elbaz}, {Bourne}, {Bruce}, {Buitrago}, {Castellano}, {Schreiber}, {Grazian},
  {McLure}, {Okumura}, {Shu}, {Wang}, {Amor{\'\i}n}, {Boutsia}, {Cappelluti},
  {Comastri}, {Derriere}, {Faber}, \& {Santini}}]{merl16}
---. 2016, {T-PHOT: PSF-matched, prior-based, multiwavelength extragalactic
  deconfusion photometry}, Astrophysics Source Code Library, , , ascl:1609.001

\bibitem[{{Navarro} {et~al.}(1997){Navarro}, {Frenk}, \& {White}}]{NFW}
{Navarro}, J.~F., {Frenk}, C.~S., \& {White}, S. D.~M. 1997, \apj, 490, 493

\bibitem[{{Oemler} {et~al.}(2017){Oemler}, {Clardy}, {Kelson}, {Walth}, \&
  {Villanueva}}]{Oemler2017}
{Oemler}, A., {Clardy}, K., {Kelson}, D., {Walth}, G., \& {Villanueva}, E.
  2017, {COSMOS: Carnegie Observatories System for MultiObject Spectroscopy},
  Astrophysics Source Code Library, , , ascl:1705.001

\bibitem[{{Oesch} {et~al.}(2018){Oesch}, {Bouwens}, {Illingworth}, {Labb{\'e}},
  \& {Stefanon}}]{Oesch2018}
{Oesch}, P.~A., {Bouwens}, R.~J., {Illingworth}, G.~D., {Labb{\'e}}, I., \&
  {Stefanon}, M. 2018, \apj, 855, 105

\bibitem[{{Oguri}(2010)}]{Oguri2010}
{Oguri}, M. 2010, \pasj, 62, 1017

\bibitem[{{Oke}(1974)}]{Oke1974}
{Oke}, J.~B. 1974, \apjs, 27, 21

\bibitem[{{Pandge} {et~al.}(2018){Pandge}, {Monteiro-Oliveira}, {Bagchi},
  {Simionescu}, {Limousin}, \& {Raychaudhury}}]{Pandge2018}
{Pandge}, M.~B., {Monteiro-Oliveira}, R., {Bagchi}, J., {et~al.} 2018, \mnras,
  2802

\bibitem[{{Parekh} {et~al.}(2017){Parekh}, {Dwarakanath}, {Kale}, \&
  {Intema}}]{Parekh2017}
{Parekh}, V., {Dwarakanath}, K.~S., {Kale}, R., \& {Intema}, H. 2017, \mnras,
  464, 2752

\bibitem[{{Paterno-Mahler} {et~al.}(2018){Paterno-Mahler}, {Sharon}, {Coe},
  {Mahler}, {Cerny}, {Johnson}, {Schrabback}, {Andrade-Santos}, {Avila},
  {Bradac}, {Bradley}, {Carrasco}, {Czakon}, {Dawson}, {Frye}, {Hoag}, {Huang},
  {Jones}, {Lam}, {Livermore}, {Lovisari}, {Mainali}, {Oesch}, {Ogaz}, {Past},
  {Peterson}, {Ryan}, {Salmon}, {Sendra-Server}, {Stark}, {Umetsu}, {Vulcani},
  \& {Zitrin}}]{Paterno-Mahler2018}
{Paterno-Mahler}, R., {Sharon}, K., {Coe}, D., {et~al.} 2018, ArXiv e-prints,
  arXiv:1805.09834

\bibitem[{{Planck Collaboration} {et~al.}(2011){Planck Collaboration}, {Ade},
  {Aghanim}, {Arnaud}, {Ashdown}, {Aumont}, {Baccigalupi}, {Balbi}, {Banday},
  {Barreiro}, \& et~al.}]{Planck2011VII}
{Planck Collaboration}, {Ade}, P.~A.~R., {Aghanim}, N., {et~al.} 2011, \aap,
  536, A7

\bibitem[{{Planck Collaboration} {et~al.}(2016{\natexlab{a}}){Planck
  Collaboration}, {Adam}, {Aghanim}, {Ashdown}, {Aumont}, {Baccigalupi},
  {Ballardini}, {Banday}, {Barreiro}, {Bartolo}, {Basak}, {Battye}, {Benabed},
  {Bernard}, {Bersanelli}, {Bielewicz}, {Bock}, {Bonaldi}, {Bonavera}, {Bond},
  {Borrill}, {Bouchet}, {Boulanger}, {Bucher}, {Burigana}, {Calabrese},
  {Cardoso}, {Carron}, {Chiang}, {Colombo}, {Combet}, {Comis}, {Couchot},
  {Coulais}, {Crill}, {Curto}, {Cuttaia}, {Davis}, {de Bernardis}, {de Rosa},
  {de Zotti}, {Delabrouille}, {Di Valentino}, {Dickinson}, {Diego}, {Dor{\'e}},
  {Douspis}, {Ducout}, {Dupac}, {Elsner}, {En{\ss}lin}, {Eriksen}, {Falgarone},
  {Fantaye}, {Finelli}, {Forastieri}, {Frailis}, {Fraisse}, {Franceschi},
  {Frolov}, {Galeotta}, {Galli}, {Ganga}, {G{\'e}nova-Santos}, {Gerbino},
  {Ghosh}, {Gonz{\'a}lez-Nuevo}, {G{\'o}rski}, {Gruppuso}, {Gudmundsson},
  {Hansen}, {Helou}, {Henrot-Versill{\'e}}, {Herranz}, {Hivon}, {Huang},
  {Ili{\'c}}, {Jaffe}, {Jones}, {Keih{\"a}nen}, {Keskitalo}, {Kisner}, {Knox},
  {Krachmalnicoff}, {Kunz}, {Kurki-Suonio}, {Lagache}, {L{\"a}hteenm{\"a}ki},
  {Lamarre}, {Langer}, {Lasenby}, {Lattanzi}, {Lawrence}, {Le Jeune},
  {Levrier}, {Lewis}, {Liguori}, {Lilje}, {L{\'o}pez-Caniego}, {Ma},
  {Mac{\'{\i}}as-P{\'e}rez}, {Maggio}, {Mangilli}, {Maris}, {Martin},
  {Mart{\'{\i}}nez-Gonz{\'a}lez}, {Matarrese}, {Mauri}, {McEwen}, {Meinhold},
  {Melchiorri}, {Mennella}, {Migliaccio}, {Miville-Desch{\^e}nes}, {Molinari},
  {Moneti}, {Montier}, {Morgante}, {Moss}, {Naselsky}, {Natoli}, {Oxborrow},
  {Pagano}, {Paoletti}, {Partridge}, {Patanchon}, {Patrizii}, {Perdereau},
  {Perotto}, {Pettorino}, {Piacentini}, {Plaszczynski}, {Polastri}, {Polenta},
  {Puget}, {Rachen}, {Racine}, {Reinecke}, {Remazeilles}, {Renzi}, {Rocha},
  {Rossetti}, {Roudier}, {Rubi{\~n}o-Mart{\'{\i}}n}, {Ruiz-Granados},
  {Salvati}, {Sandri}, {Savelainen}, {Scott}, {Sirri}, {Sunyaev}, {Suur-Uski},
  {Tauber}, {Tenti}, {Toffolatti}, {Tomasi}, {Tristram}, {Trombetti},
  {Valiviita}, {Van Tent}, {Vielva}, {Villa}, {Vittorio}, {Wandelt}, {Wehus},
  {White}, {Zacchei}, \& {Zonca}}]{Planck2015XLVII}
{Planck Collaboration}, {Adam}, R., {Aghanim}, N., {et~al.} 2016{\natexlab{a}},
  \aap, 596, A108

\bibitem[{{Planck Collaboration} {et~al.}(2016{\natexlab{b}}){Planck
  Collaboration}, {Ade}, {Aghanim}, {Arnaud}, {Ashdown}, {Aumont},
  {Baccigalupi}, {Banday}, {Barreiro}, {Barrena}, \& et~al.}]{Planck2015XXVII}
{Planck Collaboration}, {Ade}, P.~A.~R., {Aghanim}, N., {et~al.}
  2016{\natexlab{b}}, \aap, 594, A27

\bibitem[{{Postman} {et~al.}(2012){Postman}, {Coe}, {Ben{\'{\i}}tez},
  {Bradley}, {Broadhurst}, {Donahue}, {Ford}, {Graur}, {Graves}, {Jouvel},
  {Koekemoer}, {Lemze}, {Medezinski}, {Molino}, {Moustakas}, {Ogaz}, {Riess},
  {Rodney}, {Rosati}, {Umetsu}, {Zheng}, {Zitrin}, {Bartelmann}, {Bouwens},
  {Czakon}, {Golwala}, {Host}, {Infante}, {Jha}, {Jimenez-Teja}, {Kelson},
  {Lahav}, {Lazkoz}, {Maoz}, {McCully}, {Melchior}, {Meneghetti}, {Merten},
  {Moustakas}, {Nonino}, {Patel}, {Reg{\"o}s}, {Sayers}, {Seitz}, \& {Van der
  Wel}}]{Postman2012}
{Postman}, M., {Coe}, D., {Ben{\'{\i}}tez}, N., {et~al.} 2012, \apjs, 199, 25

\bibitem[{{Remolina Gonz{\'a}lez} {et~al.}(2018){Remolina Gonz{\'a}lez},
  {Sharon}, \& {Mahler}}]{Remolina2018}
{Remolina Gonz{\'a}lez}, J.~D., {Sharon}, K., \& {Mahler}, G. 2018, \apj, 863,
  60

\bibitem[{{Robertson} {et~al.}(2014){Robertson}, {Ellis}, {Dunlop}, {McLure},
  {Stark}, \& {McLeod}}]{Robertson14}
{Robertson}, B.~E., {Ellis}, R.~S., {Dunlop}, J.~S., {et~al.} 2014, \apjl, 796,
  L27

\bibitem[{{Salmon} {et~al.}(2017){Salmon}, {Coe}, {Bradley}, {Bouwens},
  {Bradac}, {Huang}, {Oesch}, {Stark}, {Sharon}, {Trenti}, {Avila}, {Ogaz},
  {Andrade-Santos}, {Carrasco}, {Cerny}, {Dawson}, {Frye}, {Hoag}, {Johnson},
  {Jones}, {Lam}, {Lovisari}, {Mainali}, {Past}, {Paterno-Mahler}, {Peterson},
  {Riess}, {Rodney}, {Ryan}, {Sendra-Server}, {Strolger}, {Umetsu}, {Vulcani},
  \& {Zitrin}}]{salmon17}
{Salmon}, B., {Coe}, D., {Bradley}, L., {et~al.} 2017, ArXiv e-prints,
  arXiv:1710.08930

\bibitem[{{Salmon} {et~al.}(2018){Salmon}, {Coe}, {Bradley}, {Brada{\v c}},
  {Strait}, {Paterno-Mahler}, {Huang}, {Oesch}, {Zitrin}, {Acebron}, {Cibirka},
  {Kikuchihara}, {Oguri}, {Brammer}, {Sharon}, {Trenti}, {Avila}, {Ogaz},
  {Andrade-Santos}, {Carrasco}, {Cerny}, {Dawson}, {Frye}, {Hoag}, {Jones},
  {Mainali}, {Ouchi}, {Rodney}, {Stark}, \& {Umetsu}}]{Salmon2018}
---. 2018, \apjl, 864, L22

\bibitem[{{Sandhu} {et~al.}(2018){Sandhu}, {Malu}, {Raja}, \&
  {Datta}}]{Sandhu2018}
{Sandhu}, P., {Malu}, S., {Raja}, R., \& {Datta}, A. 2018, \apss, 363, 133

\bibitem[{{Schwarz}(1978)}]{Schwarz1978}
{Schwarz}, G. 1978, Annals of Statistics, 6, 461

\bibitem[{{Trenti} \& {Stiavelli}(2008)}]{Trenti08}
{Trenti}, M., \& {Stiavelli}, M. 2008, \apj, 676, 767

\bibitem[{{Voges} {et~al.}(1999){Voges}, {Aschenbach}, {Boller},
  {Br{\"a}uninger}, {Briel}, {Burkert}, {Dennerl}, {Englhauser}, {Gruber},
  {Haberl}, {Hartner}, {Hasinger}, {K{\"u}rster}, {Pfeffermann}, {Pietsch},
  {Predehl}, {Rosso}, {Schmitt}, {Tr{\"u}mper}, \& {Zimmermann}}]{Voges1999}
{Voges}, W., {Aschenbach}, B., {Boller}, T., {et~al.} 1999, \aap, 349, 389

\bibitem[{{Yue} {et~al.}(2017){Yue}, {Castellano}, {Ferrara}, {Fontana},
  {Merlin}, {Amor{\'\i}n}, {Grazian}, {M{\'a }rmol-Queralto}, {Micha{\l}owski},
  {Mortlock}, {Paris}, {Parsa}, {Pilo}, {Santini}, \& {Di
  Criscienzo}}]{Yue2017}
{Yue}, B., {Castellano}, M., {Ferrara}, A., {et~al.} 2017, ArXiv e-prints,
  arXiv:1711.05130

\bibitem[{{Zheng} {et~al.}(2012){Zheng}, {Postman}, {Zitrin}, {Moustakas},
  {Shu}, {Jouvel}, {H{\o}st}, {Molino}, {Bradley}, {Coe}, {Moustakas},
  {Carrasco}, {Ford}, {Ben{\'{\i}}tez}, {Lauer}, {Seitz}, {Bouwens},
  {Koekemoer}, {Medezinski}, {Bartelmann}, {Broadhurst}, {Donahue}, {Grillo},
  {Infante}, {Jha}, {Kelson}, {Lahav}, {Lemze}, {Melchior}, {Meneghetti},
  {Merten}, {Nonino}, {Ogaz}, {Rosati}, {Umetsu}, \& {van der Wel}}]{Zheng2012}
{Zheng}, W., {Postman}, M., {Zitrin}, A., {et~al.} 2012, \nat, 489, 406

\bibitem[{{Zitrin} {et~al.}(2015){Zitrin}, {Fabris}, {Merten}, {Melchior},
  {Meneghetti}, {Koekemoer}, {Coe}, {Maturi}, {Bartelmann}, {Postman},
  {Umetsu}, {Seidel}, {Sendra}, {Broadhurst}, {Balestra}, {Biviano}, {Grillo},
  {Mercurio}, {Nonino}, {Rosati}, {Bradley}, {Carrasco}, {Donahue}, {Ford},
  {Frye}, \& {Moustakas}}]{Zitrin2015}
{Zitrin}, A., {Fabris}, A., {Merten}, J., {et~al.} 2015, \apj, 801, 44

\end{thebibliography}

\appendix
\section{candidate multiple imaeges}
\label{sect:candidate_mul}
We provide a list of candidate multiple images that were discovered in this work. These galaxies were not deemed reliable enough to be used as constraints. If confirmed with deeper observations, they could become useful lensing evidence to constrain areas in the field that are currently under-constrained. Table~\ref{tab:candidates} lists the candidate IDs and coordinates. They are plotted in Figure~\ref{fig:img-mul}.

\begin{deluxetable}{lll} 
\tablecolumns{6} 
\tablecaption{List of lensed galaxy candidates \label{tab:candidates}} 
\tablehead{\colhead{ID} & 
            \colhead{R.A.}    &  
            \colhead{Decl.}          \\[-8pt]
            \colhead{} & 
            \colhead{J2000}     &  
            \colhead{J2000}     
                   }
\startdata 
c18.1 & 64.40084583 & -11.91028778  \\ 
c18.2 & 64.40074167 & -11.91053000  \\
\hline
c19.1 & 64.38539166 & -11.90097528  \\ 
c19.2 & 64.38445834 & -11.90156944  \\ 
\hline
c20.1 & 64.39867042 & -11.91895096  \\ 
\hline
c21.1 & 64.40059584 & -11.91285222  \\ 
c21.2 & 64.39874584 & -11.91455333  \\ 
\hline
c22.1 & 64.39631667 & -11.91718722  \\ 
c22.2 & 64.39631667 & -11.91718722  \\ 
\hline
c23.1 & 64.38672916 & -11.90686278  \\ 
c23.2 & 64.38670763 & -11.90698944  \\ 
\hline
c24.1 & 64.39396249 & -11.91067667  \\ 
c24.2 & 64.39380000 & -11.91072361  \\ 
\enddata 
 \tablecomments{\textbf{Our candidate system 22 was also reported as a SL features in \cite{Pandge2018}}}
\end{deluxetable} 

\section{EASY Photo-$z$ estimates}
\label{apdx:photo-z-model-z}
\begin{figure*}
\begin{center}
\includegraphics[trim={4cm 1cm 2cm 2cm},width=\textwidth]{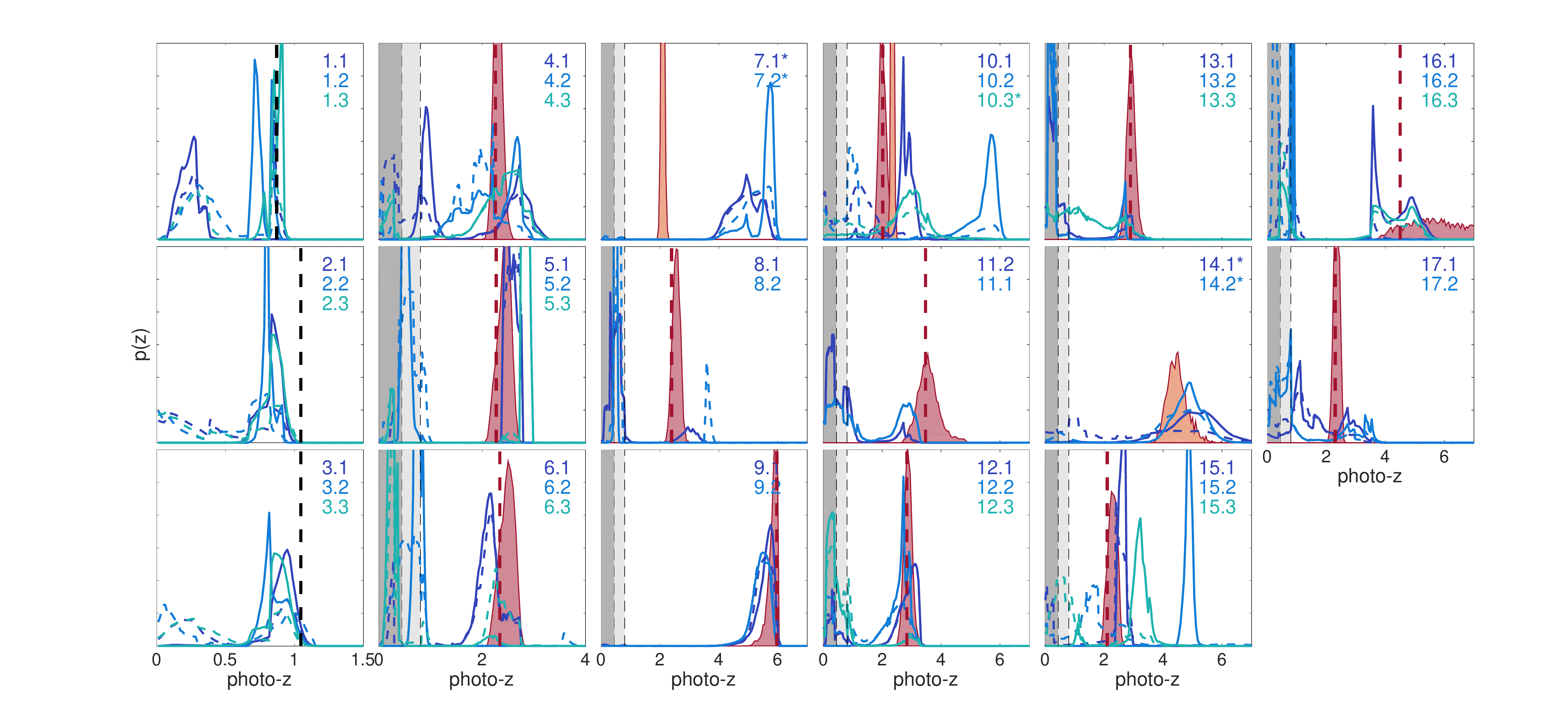}
\caption{
Same as Figure~\ref{fig:modelz-photoz}, but here both the \hst-only and \hst+\spitzer\ PDFs are computed with the same algorithm, EAZY. While there are some differences between the EASY and BPZ outputs, choosing one algorithm over the other does not change the results of this paper.  
}
\label{fig:modelz-photoz-EASY}
\end{center}
\end{figure*}

The \photoz\ estimates that were used in this analysis are computed with two different algorithms, the \hst-only analysis was done with BPZ and matches the catalogs that are publicly available on MAST. The \hst+\spitzer\ analysis uses EAZY. In this appendix, we repeat the comparison between model-predicted redshift PDFs and those of the \photoz\ estimates, using EASY for both sets of \photoz\ measurements (see Figure~\ref{fig:modelz-photoz-EASY}).
The same \hst\ photometric catalogs, also available on MAST, were used in all cases.   We find that the choice of \photoz\ algorithm does not significantly change our conclusion that the photometric redshifts and model redshifts are generally in good agreement.

\end{document}